\documentstyle[12pt,epsfig]{article}
\renewcommand{\theequation}{\thesection.\arabic{equation}}
\newcommand{\beq}{\begin{equation}}
\newcommand{\eeq}{\end{equation}}
\newcommand{\bea}{\begin{eqnarray}}
\newcommand{\eea}{\end{eqnarray}}
\newcommand{\gsim}{\raisebox{-0.07cm}{$\:\stackrel{>}{{\scriptstyle
 \sim}}\: $} }
\newcommand{\lsim}{\raisebox{-0.07cm}{$\:\stackrel{<}{{\scriptstyle
 \sim}}\: $} }
\newcommand{\ra}{\!\rightarrow\!}
\newcommand\MSb{$\overline{\mbox{MS}}$}
\newcommand{\lam}{\lambda}
\newcommand{\GE}{\gamma_{e}}
\newcommand{\DD}{{\cal D}}

\begin{document}
\setlength{\parskip}{0.32cm}
\setlength{\baselineskip}{0.57cm}
\begin{titlepage}
\noindent
{\tt hep-ph/0103123} \hfill INLO-PUB 12/00 \\
\hspace*{\fill} February 2001 \\
\vspace{1.8cm}
\begin{center}
\Large
{\bf Non-Singlet Structure Functions} \\
\vspace{0.15cm}
{\bf Beyond the Next-to-Next-to Leading Order} \\
\vspace{2.7cm}
\large
W.L. van Neerven and A. Vogt \\
\vspace{0.8cm}
\normalsize
{\it Instituut-Lorentz, University of Leiden \\
\vspace{0.1cm}
P.O. Box 9506, 2300 RA Leiden, The Netherlands} \\
\vspace{4.5cm}
{\bf Abstract}
\vspace{-0.3cm}
\end{center}
We study the evolution of the flavour non-singlet deep-inelastic
structure functions $F_{2,\rm NS}$ and $F_3$ at the next-to-next-to-%
next-to-leading order (N$^3$LO) of massless perturbative QCD. The 
present information on the corresponding three-loop coefficient 
functions is used to derive approximate expressions of these quantities 
which prove completely sufficient for values $x\! >\! 10^{-2}$ of the 
Bjorken variable. The inclusion of the N$^3$LO corrections reduces the 
theoretical uncertainty of $\alpha_s$ determinations from non-singlet 
scaling violations arising from the truncation of the perturbation 
series to less than 1\%.
We also study the predictions of the soft-gluon resummation, of 
renormalization-scheme optimizations by the principle of minimal 
sensitivity (PMS) and the effective charge (ECH) method, and of the 
Pad\'e summation for the structure-function evolution kernels. The PMS,
ECH and Pad\'e approaches are found to facilitate a reliable estimate 
of the corrections beyond N$^3$LO. 
\vfill
\end{titlepage}
%
%
\section{Introduction}
%
%
Structure functions in inclusive deep-inelastic lepton-nucleon 
scattering (DIS) are among the observables best suited for precise 
determinations of the strong coupling constant~$\alpha_s$. At present 
their experimental uncertainties result in an error $\Delta_{\rm exp\,} 
\alpha_s(M_Z^2) \simeq 0.002$ at the mass of the \mbox{$Z$}-boson 
\cite{PDG}. A further reduction of this error can be expected, 
especially from measurements at the electron-proton collider HERA after 
the forthcoming luminosity upgrade.
The standard next-to-leading order (NLO) approximation of perturbative 
QCD summarized in ref.~\cite{FP82}, on the other hand, leads to a 
theoretical error $\Delta_{\rm th\,} \alpha_s (M_Z^2) \approx 0.005$. 
This error is dominated by the uncertainty due to the truncation of the 
perturbation series as estimated from the renormalization-scale 
dependence. Hence calculations beyond NLO are required to make full use
of the present and forthcoming data on structure functions.

The ingredients necessary for next-to-next-to-leading order (NNLO)
analyses of the structure functions in Bjorken-$x$ space\footnote
 {See refs.~\cite{KKPS,SaYn} for NNLO analyses based on the 
 integer Mellin-$N$ results of refs.~\cite{moms,mnew} only.}
have not been completed up to now: Unlike the two-loop coefficient 
functions which were calculated some time ago \cite{ZvN} (and 
completely checked recently~\cite{MV1}), only partial results 
\cite{moms,mnew,lNf,lowx} have been obtained for the three-loop 
splitting functions so far. 
However, we have recently demonstrated \cite{NV1,NV2,NV3} that the 
uncertainties resulting from the incompleteness of this information are 
entirely negligible at $x\! >\! 0.05$. Moreover, these uncertainties 
are small even at much lower $x$, down to $x \simeq 10^{-4}$ at not too 
small scales, $Q^2 \gsim 10 \mbox{ GeV}^2$~\cite{NV3}.
Thus analyses of structure functions in DIS (and of total cross sections
for Drell-Yan lepton-pair production \cite{c2DY}) can be promoted to 
NNLO over a wide kinematic region. Besides more accurate determinations 
of the parton densities, such analyses facilitate a considerably 
improved theoretical accuracy $\Delta_{\rm th\,} \alpha_s(M_Z^2) \simeq 
0.002$ of the determinations of the strong coupling.

In the present article we extend, for $x\! >\! 10^{-2}$, our treatment 
\cite{NV1} of the flavour non-singlet (NS) sector dominating $\alpha_s
$-extractions from DIS to the next-to-next-to-next-to-leading order
(N$^3$LO). This extension is facilitated by two circumstances: 
The first is the existence of constraints on the three-loop coefficient 
functions which prove to be sufficiently restrictive in this region of
$x$. The seven lowest even-integer and odd-integer moments have been 
computed \cite{moms,mnew} for the structure function $F_{2,\rm NS}$ 
in electromagnetic DIS and $F_3^{\,\nu + \bar{\nu}}$ in charged-current
DIS, respectively. Furthermore the four leading large-$x$ terms of 
these functions are known from the soft-gluon resummation 
\cite{sglue,av99}.
The second circumstance is the rapid convergence of the splitting-%
function expansion in the usual \MSb\ factorization scheme also employed
in refs.~\cite{moms,mnew}. Already the impact of the three-loop 
splitting functions is small at $x\! > \! 10^{-2}$, in absolute size 
(less than 1\% on $\alpha_s(M_Z^2)\,$) as well as compared to the 
two-loop coefficient functions \cite{NV1,NV2}. Hence one can safely 
expect that the effect of the unknown four-loop splitting functions on 
determinations of $\alpha_s(M_Z^2)$ will be well below the 1\% accuracy 
we are aiming at.

As demonstrated below, the N$^3$LO approximation suffices for achieving 
this accuracy in the region $x\,\lsim\, 0.75$ usually covered by 
analyses of DIS data \cite{PDG}. Terms beyond this order are relevant 
at $x\,\gsim\, 0.8$, on the other hand, mainly due to the presence of 
large soft-gluon logarithms up to $[\,\ln^{2l-1}(1\!-\!x)\,]/(1\!-\!x)$
in the $l$-loop coefficient functions. The resummation of these 
logarithms \cite{sglue,av99} has been extended to the next-to-next-to-%
leading logarithmic (NNLL) accuracy recently \cite{av00}. Here we will 
study the predictions of this resummation for the factorization-scheme 
independent (`physical') kernels governing the scaling violations 
(`evolution') of the non-singlet structure functions. 
Other approaches to estimate higher-order corrections to these kernels, 
not restricted to very large $x$, include Pad\'e summations of the 
perturbation series \cite{Pade} as well as renormalization scheme 
optimizations such as the principle of minimal sensitivity (PMS)
\cite{PMS} and the effective charge (ECH) method~\cite{ECH}. We will 
also compare these estimates to the full NNLO and N$^3$LO evolution
kernels, and investigate the resulting predictions at order 
$\alpha_s^5$ (N$^4$LO) and beyond.

The outline of this article is as follows: In section 2 we express the
physical evolution kernels, up to N$^4$LO and NNLL accuracy, in terms 
of the corresponding splitting functions and coefficient functions. The 
information on the three-loop coefficient functions for $F_{2,\rm NS}$ 
and $F_3$ discussed above is employed in section 3 to derive 
approximate expressions for their $x$-dependence. Besides these 
functions the N$^3$LO evolution kernels also involve convolutions of 
lower-order coefficient functions for which we provide compact 
expressions in section 4. 
These results are put together in section 5 to study the effects of the 
N$^3$LO terms on the evolution of the structure functions and on the
resulting determinations of the strong coupling constant. In section 6 
we discuss the predictions of the soft-gluon resummation and of the 
Pad\'e, PMS and ECH approximations. Finally our results are summarized 
in section 7. Some relations for the convolutions in section 4 and for 
the Pad\'e approximations in section 6 can be found in the appendix.
 
\vspace*{\fill}
%
%
\newpage
\section{Fixed-order and resummed evolution kernels}
%
%
For the choice $\mu_r^2 = \mu_f^2 = Q^2$ of the renormalization and
mass-factorization scales, the structure functions 
\beq
\label{eq21a}
  {\cal F}_1 = 2F_{1,\rm NS}\:\: , \quad 
  {\cal F}_2 = \frac{1}{x} F_{2,\rm NS}\:\: , \quad
  {\cal F}_3 = F_{3}^{\,\nu \pm \bar{\nu}} 
\eeq
are in perturbative QCD given by
\bea
\label{eq21}
  {\cal F}_a(x,Q^2) 
  &\, =\, & \Big[ C_a(Q^2) \otimes q_{a,\rm NS}^{\,}(Q^2) \Big] (x) 
  \nonumber \\[2mm] &\, =\, &
      \sum_{l=0} a_s^{\, l}(Q^2) \:\Big[ c_{a,l} \otimes 
      q_{a,\rm NS}^{\,}(Q^2) \Big] (x) 
  \nonumber \\
  &\, =\, & \sum_{l=0} a_s^{\, l}(Q^2) \int_x^1 \frac{dy}{y}\, 
      c_{a,l}(y)\: q_{a,\rm NS} \bigg(\frac{x}{y},Q^2\bigg) \:\: .
\eea
Here $c_{a,l}$ represents the $l$-loop non-singlet coefficient 
functions with \mbox{$c_{a,0}(x)=\delta(1\!-\!x)$}, and $q_{a,\rm NS}$ 
stands for the respective combinations of the quark densities. The 
scale dependence of the running coupling of QCD, in this article 
normalized as 
\bea
\label{eq22a}
  a_s \:\equiv\: \frac{\alpha_s}{4\pi} \:\: , 
\eea
is governed by 
\beq
\label{eq22}
  \frac{d a_s}{d \ln Q^2} \: = \: \beta(a_s) \: = \:
  - \sum_{l=0} a_s^{\, l+2} \beta_l \:\: .
\eeq
Besides $\beta_0$ and $\beta_1$ \cite{FP82} also the coefficients 
$\beta_2$ and $\beta_3$ have been computed \cite{beta2,beta3} in the 
\MSb\ renormalization scheme adopted throughout this study. All these 
four coefficients, 
\bea
\label{eq23}
  \beta_0 &\, =\, & \: 11\: - \: 2/3\: N_f 
  \nonumber \\
  \beta_1 &\, =\, & 102 - 38/3\, N_f 
  \nonumber \\
  \beta_2 &\, =\, & \, 2857/2\, - 5033/18\, N_f + \,325/54\: N_f^2 
  \nonumber \\
  \beta_3 &\, =\,  & 29243.0 - \: 6946.30\: N_f + 405.089\, N_f^2 
                  + 1093/729\, N_f^3 \:\: ,
\eea
are required for N$^3$LO calculations. The irrational coefficients in 
Eq.\ (\ref{eq23}) and in Eqs.\ (\ref{eq212}) and (\ref{eq215}) below
have been truncated to six digits for brevity. $N_f$ denotes the 
number of effectively massless flavours (mass effects are not 
considered in this article). Finally the evolution 
equations for the parton densities in Eq.~(\ref{eq21}) read
\bea
\label{eq24}
  \frac{d}{d \ln Q^2} \, q_{a,\rm NS}^{\,}(x, Q^2) 
  &\, =\, & \Big[ P_a(Q^2) \otimes q_{a,\rm NS}^{\,}(Q^2)\Big] (x)
  \nonumber \\
  &\, =\, & \sum_{l=0} a_s^{\, l+1} \Big[ P_{a,l}\otimes 
        q_{a,\rm NS}^{\,}(Q^2)\Big] (x) \:\: ,
\eea
where $\otimes$ abbreviates the Mellin convolution written out in the
third line of Eq.~(\ref{eq21}). Like the coefficient functions 
$c_{a,l\,}(x)$, the $(l\! +\! 1)$-loop splitting functions $P_{a,l}(x)$ 
are scale independent for the above choice of $\mu_r$ and $\mu_f$.

Explicit expressions up to order $\alpha_s^2$ can be found in refs.\
\cite{CFP} and \cite{ZvN} for the non-singlet splitting functions and
coefficient functions, respectively. For the third-order splitting
functions $P_{a,2}(x)$ we will employ our approximate expressions of 
ref.~\cite{NV3}. The three-loop coefficient functions $c_{a,3}(x)$
are the subject of section 3 below.

It is convenient to express the scaling violations of the non-singlet 
structure functions in terms of these structure functions themselves, 
thus explicitly eliminating any dependence on the factorization scheme 
and the scale $\mu_f$. The corresponding `physical' evolution kernels 
$K_a$ for $\mu_r^2 = Q^2$ can be derived by differentiating 
Eq.~(\ref{eq21}) with respect to $Q^2$ by means of Eqs.~(\ref{eq22}) 
and (\ref{eq24}), and finally eliminating $q_{a,\rm NS}^{\,}$ using the 
inverse of Eq.~(\ref{eq21}). Suppressing the dependences on $x$ and 
$Q^2$ one arrives at the evolution equations
\bea
\label{eq25}
  \frac{d}{d \ln Q^2} \, {\cal F}_{a}
  &\, =\, & \left( P_a + \beta \,\frac{d \ln C_a}{d a_s} \right) 
    \otimes {\cal F}_{a} 
  \nonumber \\[1mm]
  &\, = \, & K_{a} \otimes {\cal F}_{a} \: = \: 
    \sum_{l=0}  a_s^{\, l+1} K_{a,l} \otimes {\cal F}_{a} \nonumber\\
  &\, =\, & \left\{ a_s P_0 + \sum_{l=1}  a_s^{\, l+1} 
    \left( P_{a,l} - \sum_{k=0}^{l-1} \beta_k \tilde{c}_{a,l-k} \right) 
    \right\} \otimes {\cal F}_{a} \\[-6mm] \nonumber
\eea
with
\bea 
\label{eq26} 
  & & \nonumber\\[-8mm]
  \tilde{c}_{a,1} &\, =\, & c_{a,1} 
  \nonumber \\
  \tilde{c}_{a,2} &\, =\, & 2\, c_{a,2} - c_{a,1}^{\,\otimes 2} 
  \nonumber \\
  \tilde{c}_{a,3} &\, =\, & 3\, c_{a,3} - 3\, c_{a,2} \otimes c_{a,1} 
    + c_{a,1}^{\,\otimes 3}  
  \\
  \tilde{c}_{a,4} &\, =\, & 4\, c_{a,4} - 4\, c_{a,3} \otimes c_{a,1}
    - 2\, c_{a,2}^{\,\otimes 2} + 4\, c_{a,2} \otimes 
   c_{a,1}^{\,\otimes 2} - c_{a,1}^{\,\otimes 4} \:\: .
 \nonumber 
\eea
In Eq.~(\ref{eq26}) we have used the abbreviation $f^{\,\otimes l}$ for 
the $(l\! -\! 1)$-fold convolution of a function $f(x)$ with itself, 
i.e., $ f^{\,\otimes 2} = f \otimes f$ etc.  The generalizations 
${\cal K}_{a,l}$ of the kernels $K_{a,l}$ in Eq.~(\ref{eq25}) to 
$\mu_r^2 \neq Q^2$ can be obtained by expanding$\,$\footnote
{Up to the fifth order this expansion can be read off from the 
 $K_{a,0}$ terms of Eq.~(\ref{eq27}).}
$a_s(Q^2)$ in terms of $a_s(\mu_r^2)$ and $L =  \ln (Q^2/\mu_r^2)\,$,
yielding
%
\bea 
\label{eq27} 
 {\cal K}_{a,0} &\, =\, & K_{a,0} 
 \nonumber \\[0.7mm]
 {\cal K}_{a,1} &\, =\, & K_{a,1} - \beta_0 L K_{a,0} 
 \nonumber \\[0.8mm]
 {\cal K}_{a,2} &\, =\, & K_{a,2} - 2\beta_0 L K_{a,1} 
   - (\beta_1 L - \beta_0^2 L^2)\, K_{a,0} 
 \nonumber \\[1mm]
 {\cal K}_{a,3} &\, =\, & K_{a,3} - 3\beta_0 L K_{a,2} 
   - (2\beta_1 L - 3\beta_0^2 L^2)\, K_{a,1}
 \nonumber \\ & & \mbox{}  
   - \Big( \beta_2 L - \frac{5}{2}\beta_1\beta_0 L^2 
   + \beta_0^3 L^3 \Big)\, K_{a,0}
  \nonumber \\[1mm]
 {\cal K}_{a,4} &\, =\, & K_{a,4} - 4\beta_0 L K_{a,3}
   - (3\beta_1 L - 6\beta_0^2 L^2)\, K_{a,2}
 \nonumber \\[0.5mm] & & \mbox{}  
   - (2 \beta_2 L - 7 \beta_1 \beta_0 L^2 + 4\beta_0^3 L^3)\,
   K_{a,1}
 \\ & & \mbox{}
   - \Big( \beta_3 L - 3\beta_2\beta_0 L^2 - \frac{3}{2} \beta_1^2 
   L^2 + \frac{13}{3} \beta_1 \beta_0^2 L^3 - \beta_0^4 L^4 \Big)\,
   K_{a,0} \:\: . \nonumber
\eea
At N$_{\,}^m$LO the terms up to $l= m$ are included in Eq~(\ref{eq25}).
Hence Eqs.~(\ref{eq26}) and (\ref{eq27}) formally specify the evolution 
kernels ${\cal K}_a$ up to N$_{\,}^4$LO. Their extension to higher 
orders is straightforward but irrelevant for the time being, as at 
least the coefficient functions beyond four loops will not be calculated
in the foreseeable future.

The leading terms of the coefficient functions for $x \!\rightarrow\! 
1$, however, are known to all orders from the soft-gluon resummation 
\cite{sglue,av99,av00}. Switching to Mellin moments, 
\beq
\label{eq28} 
  f^N \: =\: \int_0^1\! dx \, x^{N-1} f(x) \:\: ,
\eeq
for the remainder of this section, the large-$N$ (large-$x$) behaviour 
of the coefficient functions in Eq.~(\ref{eq21}) takes the form
\beq
\label{eq29} 
  C_{\rm res}^N \: = \: g_0^{}(a_s) \,\exp \Big\{\ln N\, g_1^{}(\lam) 
  + g_2^{}(\lam) + a_s g_3^{}(\lam) + {\cal O}(a_s^2 f(\lam)) \Big\}
\eeq
up to terms which vanish for $N \!\rightarrow \!\infty$. Here we have
used the abbreviation
\beq
\label{eq210} 
  \lambda \: =\: a_s \beta_0 \ln N \: =\: 
  \frac{\alpha_s}{4\pi} \,\beta_0 \ln N \:\: ,
\eeq
and we have again put $\mu_r^2 = \mu_f^2 = Q^2$. By virtue of the first 
line of (\ref{eq25}), Eq.~(\ref{eq29}) leads to the following 
expression for the resummed kernel up to next-to-next-to-leading 
logarithmic (NNLL) accuracy \cite{av00}:
\bea
\label{eq211} 
  K_{\rm res}^N &\, =\, &
  \mbox{} - (A_1\,a_s + A_2\,a_s^2 + A_3\,a_s^3)\, \ln N \: - \:
  \bigg( 1+ \frac{\beta_1}{\beta_0} a_s + \frac{\beta_2}{\beta_0}
  a_s^2 \bigg)\, \lambda^2 \frac{dg_1^{}}{d\lambda}
  \nonumber \\ & & \mbox{}
  - \Big( a_s \beta_0 + a_s^2 \beta_1 \Big)\, \lambda
  \frac{dg_2^{}}{d\lambda} \: - \: a_s^2 \beta_0\,
  \frac{d}{d\lambda} \Big( \lambda g_3^{}(\lambda)\Big)
  \: + \: {\cal O}(a_s^3 (f(\lam)) \:\: . \quad
\eea
Thus the leading logarithmic (LL), next-to-leading logarithmic (NLL) 
and NNLL large-$N$ contributions to the physical evolution kernels are
of the form $(a_s \ln N)^n$, $a_s (a_s \ln N)^n$ and $a_s^2 
(a_s \ln N)^n$, respectively. This is in contrast to the coefficient
functions which receive contributions up to $(a_s \ln^2 N)^n$.
The constants $A_l$ in Eq.~(\ref{eq211}) are the coefficients of the 
leading \cite{spx1} large-$x$ terms $1/[1-x]_+$ of the $l$-loop
\MSb\ splitting functions --- recall that 
\beq
  f^N \: = \: - \ln N + {\cal O}(1) \quad \mbox{ for } \quad
  f(x)\: = \: 1/[1-x]_+ \:\: . 
\eeq
As in Eq.~(\ref{eq23}) inserting the numerical values for the QCD 
colour factors $C_A$ and $C_F$, these constants are given by
\beq
\label{eq212}
  A_1 \: = \: 16/3  \: , \:\:\:
  A_2 \: = \: 66.4732 - 160/27\, N_f^{} \:\: ,
\eeq
and the yet approximate, but sufficiently accurate three-loop result 
\cite{lNf,NV3}
\beq
\label{eq213}
  A_3 \: = \: (1178.8 \pm 11.5) - (183.95 \pm 0.85)\, N_f 
     - 64/81\, N_f^2 \:\: .
\eeq
Inserting the explicit form of the functions $g_1^{}$, $g_2^{}$ 
\cite{sglue} and $g_3^{}$ \cite{av00} in Eq.~(\ref{eq211}) and 
restoring the dependence on $L = \ln (Q^2/\mu^2_r)$ leads to
\bea
\label{eq214}
  {\cal K}_{\rm res}^N &\, = & \:\frac{A_1}{\beta_0}\, \ln (1-\lambda ) 
  \: +\: a_s L\, A_1 \,\frac{\lam}{1-\lam}
  \: +\: a_s^2 L^2\, A_1\beta_0 \,\frac{\lam (\lam - 2)}{2(1-\lam)^2}
  \nonumber \\ & & \mbox{}
  + a_s \: 
    \Big\{ A_1 \beta_1 \,\Big( \lam + \ln (1-\lam) \Big) 
    + (B_1 - A_1 \GE) \beta_0^2 \lam - A_2 \beta_0 \lam \Big\} 
    \,\frac{1}{\beta_0^2 (1-\lam)}
  \nonumber \\[1mm] & & \mbox{}
  + a_s^2 L \,\Bigg\{ 
    \Big[ (B_1 - \GE A_1) \beta_0 - A_2 \Big] \frac{\lam (\lam - 2)}
    {(1-\lam)^2} 
    - A_1 \beta_1 \, \frac{\ln (1-\lam)}{\beta_0 (1-\lam)^2} \Bigg\}
  \nonumber \\[0.5mm] & & \mbox{}
  + a_s^2 \,\Bigg\{ 
  \Big[ 
    A_1 (\GE^2 + \zeta_2) \beta_0^2 + 2 A_2 \GE \beta_0 + A_3 
    - 2 B_1 \GE \beta_0^2 - 2 B_2 \beta_0
  \Big] 
  \,\frac{\lam (\lam - 2)}{2\beta_0 (1- \lambda)^2} 
  \quad \nonumber \\[1mm] & & \mbox{} \quad \quad 
  + \Big[ \,
    2 (A_1 \GE - B_1) \beta_1 \beta_0^2 \, \ln (1- \lam)
    + A_1 \beta_1^2 \, \Big( \lam^2 - \ln^2 (1-\lam) \Big)
  \nonumber \\[1.5mm] & & \mbox{} \quad \quad \quad 
    - A_1 \beta_2 \beta_0 \, \lam^2 
    - A_2 \beta_1 \beta_0 \, \Big( \lam^2 - 2\lam - 2\ln (1-\lam) \Big)
  \Big] 
  \frac{1}{2\beta_0^3 (1- \lambda)^2} 
  \nonumber \\ & & \mbox{} \quad \quad
  + 2 D_2 \frac{\lam (1 - \lam )}{(1-2\lam)^2} \Bigg\}
  \: + \: {\cal O}(a_s^3 f(\lambda,L)) 
  \\[1mm]
  &\, \equiv & \:\: 
  \sum_{l=0}^{\infty} a_s^{l+1}\, {\cal K}_{{\rm res},l}^N \:\: .
  \nonumber 
\eea
Here $\zeta_2 = \pi^2/6$, and $\gamma_e$ represents the Euler-Mascheroni
constant, $\gamma_e \simeq 0.577216$. Furthermore $B_1 = -4$ \cite
{sglue}, and the constants $B_2$ and $D_2$ are related by~\cite{av00}
\beq
\label{eq215}
  B_2 + D_2 \: = \: 36.2657 + 6.34888\, N_f \:\: .
\eeq
We will return to the latter coefficients at the end of the next 
section.

After subtracting the terms up to order $a_s^{m+1}$ in ${\cal K}^N_{\rm
res}$ already taken into account in the N$^m$LO terms (\ref{eq27}), 
Eq.~(\ref{eq214}) can be added to these fixed-order results to obtain 
the (N$^m$LO + resummed) approximation for the non-singlet evolution 
kernels,
\beq
\label{eq216}
  {\cal K}^N_{{\rm N}^m{\rm LO} + \rm res} \: = \: \sum_{l=0}^m 
  a_s^{l+1}\, ({\cal K}_{a,l}^N - {\cal K}_{{\rm res},l}^N) 
  \, + \, {\cal K}^N_{\rm res} \:\: .
\eeq
Due to the renormalon singularities at $\lam = 1$ and $\lam = 1/2$ in 
Eq.~(\ref{eq214}) the resummed evolution equations cannot be uniquely
inverted to $x$-space, unlike the fixed-order approximations ${\cal K}
^N_{a,l}\, {\cal F}_a^N$. Note that strength of these singularities 
--- located at $N \simeq 2000$ and $N \simeq 45\,$ for $\lam = 1$ and
$\lam = 1/2$, respectively, at $\alpha_s = 0.2$ and $N_f = 4$ --- 
increases with the order of the soft-gluon expansion: the behaviour is
logarithmic at the leading-log level, but involves poles of order $k$ 
for the N$^k$LL approximations. 
For our numerical study of the all-order case at the end of section 6 
we will use the standard `minimal prescription' contour \cite{cont} for 
the Mellin inversion. This contour runs to the left of the renormalon 
singularities, but to the right of all other poles in the $N$-plane. 
%
%
%
\section{The 3-loop non-singlet coefficient functions} 
%
\setcounter{equation}{0}
The $l$-loop coefficient functions $c_{a,l}$ for the non-singlet 
structure functions ${\cal F}_{a=1,2,3}$ defined in Eq.~(\ref{eq21a}) 
can be represented as
\bea
\label{eq31}
  c^{}_{a,l}(x,N_f) &\, =\, & 
  \sum_{m=0}^{2l-1} A_{l}^{(m)}\, \DD_m \, +\,  
  \tilde{B}_l^{}\, \delta(1-x) \, +\,  c_{a,l}^{\,\rm smooth}(x,N_f)
  \nonumber \\ &\, +\, & 
  \,\sum_{n=1}^{2l-1} \Big( C_{a,l}^{(n)}\, L_1^n 
  + D_{a,l}^{(n)}\, L_0^n \Big) \:\: .
\eea
Here $A_{l}^{(m)}\!$, $\tilde{B}_l^{}$, $C_{a,l}^{(n)}$ and 
$D_{a,l}^{(n)}$ are numerical coefficients which in general depend on 
the number of flavours $N_f$, and we have employed the abbreviations 
\beq
\label{eq32}
  \DD_k \: = \:\left[ \frac{\ln^k (1-x)}{1-x} \right]_+ 
  \: ,\quad L_1 \: = \: \ln (1-x) 
  \: ,\quad L_0 \: = \: \ln x  
\eeq
for the +-distributions (see Eqs.~(\ref{eq33}) and (\ref{eq34}) below)  
and the end-point logarithms. 
The functions $c_{a,l}^{\,\rm smooth}(x,N_f)$ collect all contributions
which are finite for $0\!\leq \! x \!\leq\! 1$. This regular term
constitutes the mathematically complicated part of Eq.~(\ref{eq31}), it 
involves higher transcendental functions like the harmonic 
polylogarithms introduced in ref.~\cite{hpol}. 
As usual, the +-distributions are defined via
\beq
\label{eq33}
  \int_0^1 \! dx \, a(x)_+ f(x) \: = \: \int_0^1 \! dx \, a(x)
  \,\{ f(x) - f(1) \} 
\eeq 
where $f(x)$ is a regular function. The convolutions with the 
distributions occurring in Eq.~(\ref{eq31}) can be written as$\,$%
\footnote{The second line of Eq.~(\ref{eq34}) is given by $\, -xf(x) 
 \int_0^x \!  dy \, a(y)$ for a general +-distribution $[a(x)]_+$.}
\bea
\label{eq34}
  x[\DD_k \otimes f](x) &\, = \, & 
  \int_x^1 \! dy \: \frac{\ln^k (1-x)}{1-x} \left\{ \frac{x}{y}\, 
  f\!\left( \frac{x}{y} \right) - xf(x) \right\} 
  \nonumber \\[1mm] & & \mbox{}
  + \, xf(x) \frac{1}{k+1} \ln^{k+1}(1-x) \:\: .
\eea
As already indicated in Eq.~(\ref{eq31}), the coefficients of 
$\DD_m$ and of $\delta(1-x)$ are independent of the choice of the
structure function. 
 
The three-loop contributions $c_{S,3}^{}$ known from the soft-gluon 
resummation read
\bea
\label{eq35}
  c_{S,3}^{}(x,3) & \, = \, &
        512/27\, (\DD_5 - L_1^5) - 14400/81\, \DD_4
      + 264.062\, {\cal D}_3 + 1781.704\, \DD_2
  \nonumber \\[1mm] 
  c_{S,3}^{}(x,4) & \, = \, &
        512/27\, (\DD_5 - L_1^5) - 13760/81\, \DD_4
      + 188.210\, \DD_3 + 1962.178\, \DD_2 
  \\[1mm] 
  c_{S,3}^{}(x,5) & \, = \, &
        512/27\, (\DD_5 - L_1^5) - 13120/81\, \DD_4
      + 113.938\, \DD_3 + 2131.195\, \DD_2 \:\: , \quad
  \nonumber
\eea
where we have again truncated the irrational coefficients and 
restricted ourselves to the practically relevant cases $N_f = 3$, 4 
and 5. Besides the $\DD_m$-terms determined in ref.~\cite{av99}, 
Eq.~(\ref{eq35}) also includes the leading integrable large-$x$ 
logarithm. The general relation between the coefficients of this term 
and the leading +-distribution has been conjectured in ref.~\cite{KLS}. 
Eqs.~(\ref{eq35}) complement the main present constraints on $c_{a,3}
^{}(x,N_f)$ provided by the computation \cite{moms,mnew} of the seven 
lowest even-integer and odd-integer moments (\ref{eq28}), respectively, 
for electromagnetic (e.m.) DIS and the charged-current (CC) combination 
$F_3^{\,\nu +\bar{\nu}}$. Note that the coefficients of the leading 
small-$x$ logarithms are presently unknown here, unlike for the 
splitting functions and the singlet coefficient functions \cite{lowx}.

We use this information for approximate reconstructions of 
$c_{2,3}^{}(x,N_f)$ and $c_{3,3}^{}(x,N_f)$ at $N_f = 3$, 4 and 5. 
Our method is analogous to the treatment of the three-loop splitting 
functions in  refs.~\cite{NV1,NV2,NV3}: A simple ansatz is chosen for
$c_{a,3}^{\,\rm smooth}$ in Eq.~(\ref{eq31}), and its free parameters 
are determined from the available moments together with a reasonably
balanced subset of the coefficients $A_{3}^{(0,1)\!}$, $\tilde{B}_3^{}$,
$C_{a,3}^{(n)}$ and $D_{a,3}^{(n)}$. The ansatz for $c_{a,3}^{\,\rm 
smooth}$ and the choice of the non-vanishing end-point parameters are 
then varied in order to estimate the residual uncertainty of 
$c_{a,3}^{}$. 
Specifically we keep $A_{3}^{(1)}$; one of each pair $A_{3}^{(0)}$ and 
$\tilde{B}_3^{}$, $C_{a,3}^{(4)}$ and $C_{a,3}^{(3)}$, $C_{a,3}^{(2)}$ 
and $C_{a,3}^{(1)}$; one or two of the $D_{a,3}^{(n < 4)}$; and one or 
two parameters of a polynomial up to second order in $x$ representing 
$c_{a,3}^{\,\rm smooth}$. For a few combinations the resulting 
system of linear equations which fixes these parameters by the seven 
moments becomes almost singular, resulting in exceptionally large 
numerical coefficients. After rejecting those about 5\% of the 
combinations for which the modulus of at least one parameter exceeds 
$10^5$, we are left with about 90 approximations for each case. 

Before we present the approximate results for $c_{a,3}^{}(x,N_f)$, it
is appropriate to illustrate our procedure by applying it to a known
quantity, for which we choose the two-loop e.m.\ coefficient function
$c_{2,2}^{}(x,N_f\! =\! 4)$. Adopting the coefficients of $\DD_3$
and $\DD_2$ defined in Eq.~(\ref{eq32}) from the soft-gluon
exponentiation, the procedure described in the preceding paragraph is
applied to this function with the small adjustment that two of the 
$C_{2,2}^{(1,2,3)}$ are kept as $C^{(4)}$ does not occur at two loops
according to Eq.~(\ref{eq31}). Also here we reject a couple of 
combinations, those with parameter(s) of modulus above $3\cdot 10^3$.
The remaining about 70 approximations are compared to the exact result
of ref.~\cite{ZvN} in Fig.~1.

\begin{figure}[htb]
\vspace*{1mm}
\centerline{\epsfig{file=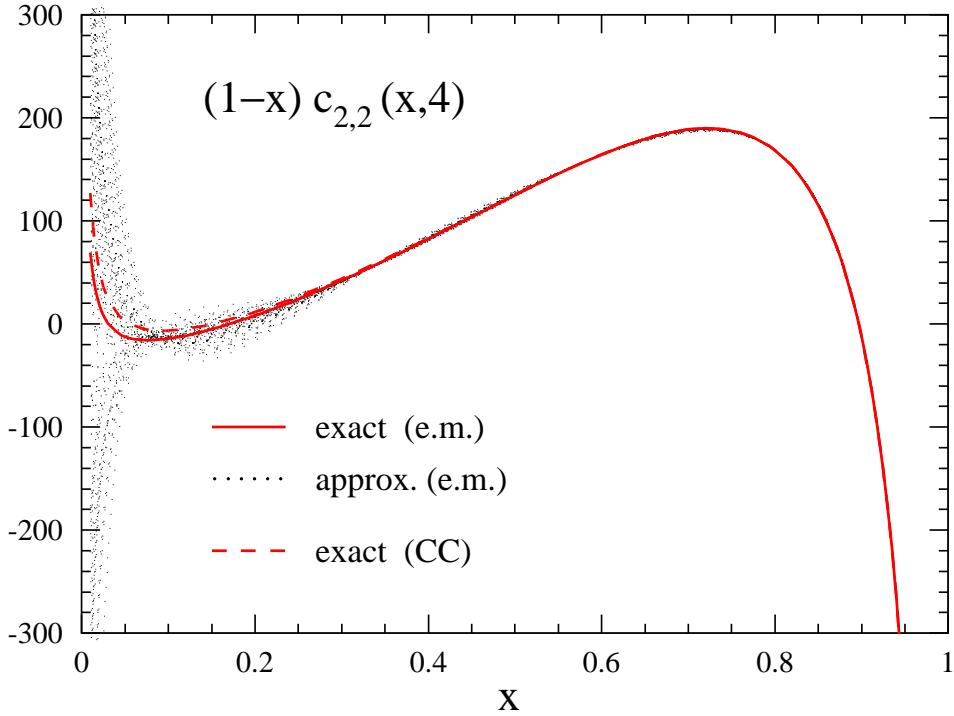,width=13cm,angle=0}}
\vspace{-1.5mm}
\caption{Approximations for the two-loop coefficient functions $c_{2,2}
 ^{}(x,N_f\! =\! 4)$ for $F_{2,\rm NS^{\,\!}}^{\,\rm e.m._{\,\!}}$ 
 obtained from the lowest seven even-integer moments and the two 
 leading soft-gluon terms, compared to the exact result of 
 ref.~\cite{ZvN}. Also shown is the corresponding exact coefficient
 function for $F_{2,\rm NS}$ in charged-current DIS.}
\vspace{1mm}
\end{figure}

The seven lowest even-integer moments supplemented by the soft-gluon
coefficients $A_{2}^{(m>1)}$ prove to constrain $c_{2,2}^{}(x)$ rather 
tightly at $x \gsim 0.3$. The region $0.1 \lsim x \lsim 0.3$ is less
accurately covered, and at $ x\! <\! 0.1$ the lack of small-$x$ 
information mentioned above becomes very prominent.
Also shown in Fig.~1 is the exact result \cite{ZvN} for $c_{2,2}^{}$
in charged-current DIS. The difference to the electromagnetic case 
--- originating in a sign difference of the contributions from $\gamma 
/ W+ q \ra q+ q+ \bar{q}$ with identical quarks in the final-state --- 
is clearly visible only at $x \lsim 0.2$. The effect of this difference
on the evolution of the structure functions at NNLO is unnoticeable at
$x \gsim 0.1$, and amounts to less that 1\% for $x > 0.01$, see Fig.~11
of ref.~\cite{NV1}. We expect that the corresponding three-loop effect
will at least not be larger. Hence the approximations for $c_{2,2}^{}
(x,N_f)$, constructed for the e.m. case, should be applicable also for 
neutrino DIS without introducing any relevant error.

\begin{figure}[p]
\vspace*{1mm}
\centerline{\epsfig{file=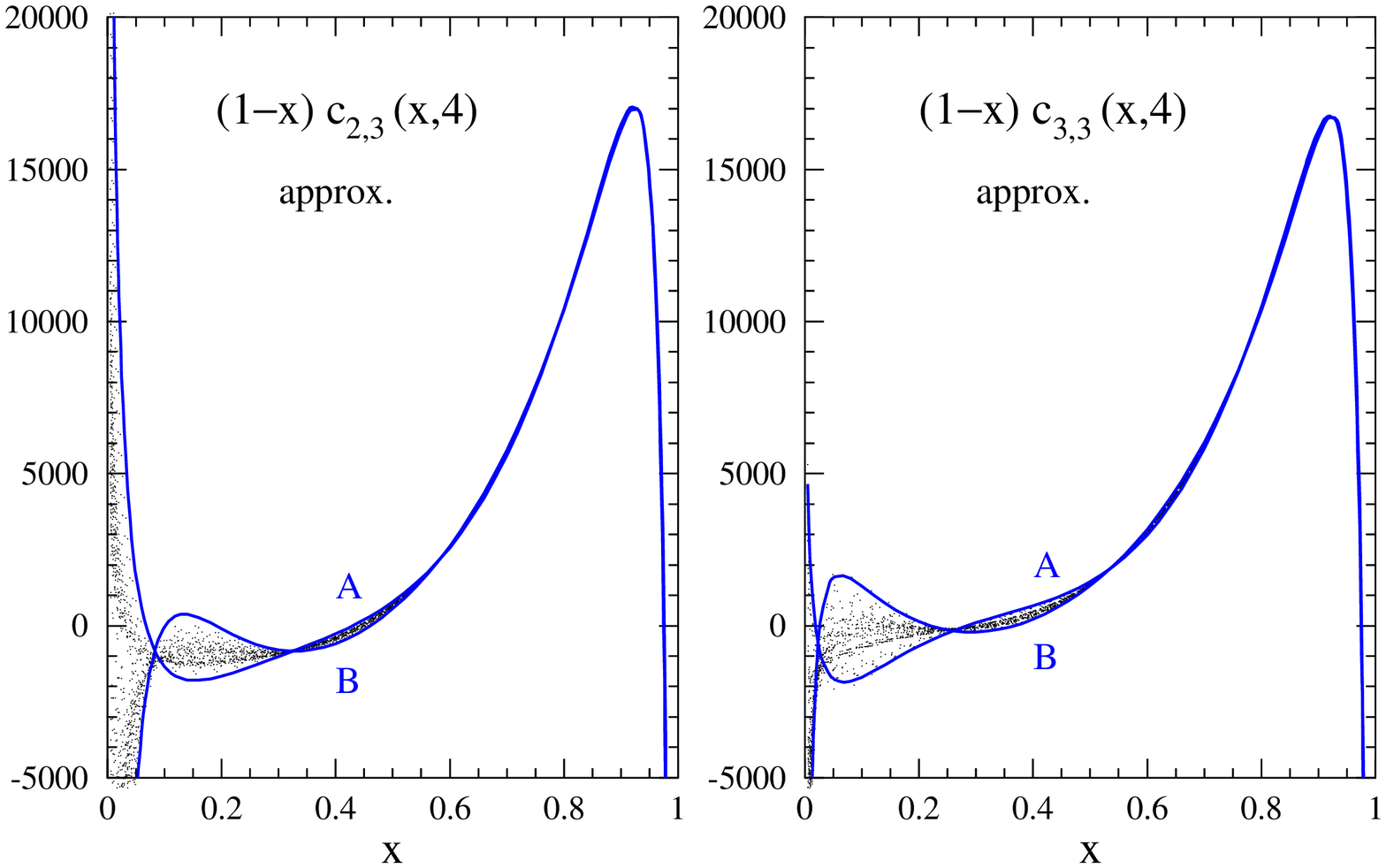,width=15cm,angle=0}}
\vspace{-1.5mm}
\caption{Approximations for the three-loop coefficient functions for
 $F_{2,\rm NS^{\,\!}}^{\,\rm e.m._{\,\!}}$ (left) and the charged-%
 current $F_{3^{\,\!}}^{\,\nu+\bar{\nu}_{\,\!}}$ (right) derived from
 the respective seven lowest moments~\cite{moms,mnew} and the
 soft-gluon terms (\ref{eq35}). The full lines show the selected
 functions (\ref{eq36}) and (\ref{eq37}).}
\vspace{1mm}
\end{figure}
\begin{figure}[p]
\vspace*{1mm}
\centerline{\epsfig{file=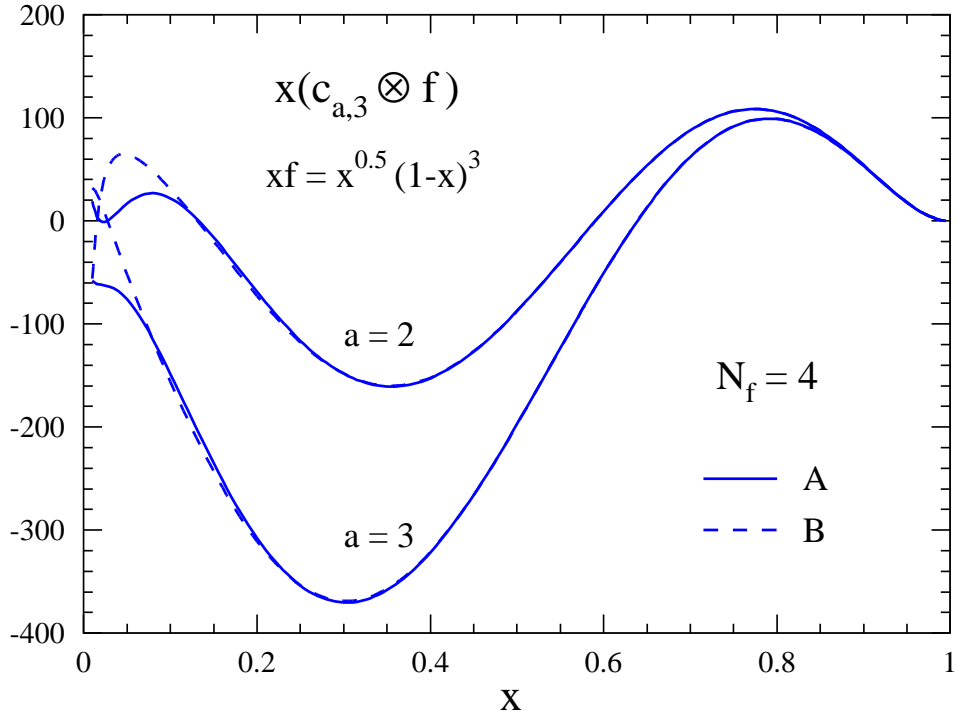,width=13cm,angle=0}}
\vspace{-1.5mm}
\caption{The convolution of the approximations (\ref{eq36}) and
 (\ref{eq37}) selected from the previous figure with a shape typical of
 hadronic non-singlet distributions.}
\vspace{1mm}
\end{figure}

The corresponding approximations for $c_{2,3}^{}$ and $c_{3,3}^{}$ are
shown in Fig.~2 for $N_f = 4$ (concerning the scale of the ordinate
recall the rather small expansion parameter (\ref{eq22a})$\,$).
As expected, the accuracy pattern is qualitatively similar to the
two-loop case of Fig.~2. The uncertainty of $c_{3,3}^{}$ is smaller
than that of $c_{2,3}^{}$ at small $x$, since for $c_{3,3}^{}$ the
lowest calculated moment, $N\! =\! 1$, is closer to the location of the
rightmost pole at $N\! =\! 0$. For both functions two representatives,
denoted by $A$ and $B$, are selected which rather completely cover the
uncertainty bands. With $c_{S,3}^{}$ of Eq.~(\ref{eq35}) these
representatives read
\bea
\label{eq36}
  c_{2,3}^{A}(x,4) &\, =\, & c_{S,3}^{}(x,4)
    - 6456.231\, {\cal D}_1 - 1085.97\, \delta(\! 1-\! x)
    + 258.876\, L_1^4
  \nonumber \\ & & \mbox{} \!\!\!
    - 22430.79\, L_1 - 74705.15\, x^2 - 4062.14\, (2-x) - 313.0\, L_0^3
  \nonumber \\[1mm]
  c_{2,3}^{B}(x,4) &\, =\, & c_{S,3}^{}(x,4)
    - 5081.227\, {\cal D}_1 + 5028.23\, {\cal D}_0 + 1059.423\, L_1^4
  \\ & & \mbox{} \!\!\!
    - 7292.84\, L_1^2 - 17741.28\, (2-x) - 18154.5\, L_0
    + 1168.02\, L_0^3
  \quad\quad \nonumber
\eea
and
\bea
\label{eq37}
  c_{3,3}^{A}(x,4) &\, =\, & c_{S,3}^{}(x,4)
    - 6940.648\, {\cal D}_1 - 1526.23\, {\cal D}_0 - 42.598\, L_1^4
  \nonumber \\ & & \mbox{} \!\!\!
    - 33562.64\, L_1 - 91639.87\, x^2 - 5898.84 + 424.49\, L_0^2
  \nonumber \\[1mm]
  c_{3,3}^{B}(x,4) &\, =\, & c_{S,3}^{}(x,4)
    - 4907.988\, {\cal D}_1 + 5587.906\, {\cal D}_0 + 1092.436\, L_1^4
  \quad\quad\\ & & \mbox{} \!\!\!
    - 8267.99\, L_1^2 - 18120.78 - 7083.63\, L_0 + 283.59\, L_0^3
  \:\: . \nonumber
\eea

The uncertainty bands of Fig.~2 do not directly indicate the range of
applicability of our approximations, as the coefficient function enter 
the structure functions and their evolution only via the smoothening 
convolution (\ref{eq21}) with non-perturbative initial distributions. 
In Fig.~3 we therefore present the convolutions of the results 
(\ref{eq36}) and (\ref{eq37}) with a typical non-singlet shape. This
illustration shows that the residual uncertainties of $c_{a,3}^{}$ do 
not lead to any relevant effects for $x \geq 0.1$. The situation at 
smaller $x$ depends on the numerical size of the $c_{a,3}^{}$ 
contributions to the evolution kernels given by Eqs.~(\ref{eq25}) and
(\ref{eq26}). Anticipating our findings in section~5, we note that 
these contributions are actually unproblematically small for 
$x > 10^{-2}$.

The results for $N_f = 3$ and $N_f = 5$ are similar to those presented
in Fig.~2 and Fig.~3. For brevity they are not displayed graphically in 
this article. The selected approximations for $N_f = 3$ are given by
\bea
\label{eq38}
  c_{2,3}^{A}(x,3) &\, =\, & c_{S,3}^{}(x,3)
    - 6973.782\, {\cal D}_1 - 3929.78\, \delta(1\! -\! x) 
    + 232.676\, L_1^4 \quad\quad
  \nonumber \\ & & \mbox{} \!\!\!
    - 35949.75\, L_1 - 93141.53\, x^2 - 10283.51 - 418.43\, L_0^3
  \nonumber \\[1mm]
  c_{2,3}^{B}(x,3) &\, =\, & c_{S,3}^{}(x,3)
    - 5575.903\, {\cal D}_1 + 5474.48\, {\cal D}_0 + 927.478\, L_1^4
  \\ & & \mbox{} \!\!\!
    - 4646.38\, L_1^2 - 23345.85 - 11094.92\, L_0 + 759.69\, L_0^3
  \nonumber
\eea
and
\bea
\label{eq39}
  c_{3,3}^{A}(x,3) &\, =\, & c_{S,3}^{}(x,3)
    - 7018.496\, {\cal D}_1 - 2957.33\, \delta(1\! -\! x) 
    + 162.667\, L_1^4
  \nonumber \\ & & \mbox{} \!\!\!
    - 28363.91\, L_1 - 74640.87\, x^2 - 5720.48\, (1+x) + 330.21\, L_0^2
  \quad\quad  \nonumber \\[1mm]
  c_{3,3}^{B}(x,3) &\, =\, & c_{S,3}^{}(x,3)
    - 5160.335\, {\cal D}_1 + 6896.360\, {\cal D}_0 + 1109.512\, L_1^4
  \\ & & \mbox{} \!\!\!
    - 7715.43\, L_1^2 - 20541.22 - 7595.83\, L_0 + 290.34\, L_0^3
  \:\: . \nonumber
\eea
The corresponding functions for $N_f = 5$ read
\bea
\label{eq310}
  c_{2,3}^{A}(x,5) &\, =\, & c_{S,3}^{}(x,5)
    - 5951.174 \, {\cal D}_1 - 391.37\, {\cal D}_0 - 2341.422\, L_1^3
  \nonumber \\ & & \mbox{} \!\!\!
    + 19986.58\, L_1 + 5517.39\, (2-x) + 5969.63\, L_0 - 284.23\, L_0^3
  \nonumber \\[1mm]
  c_{2,3}^{B}(x,5) &\, =\, & c_{S,3}^{}(x,5)
    - 4802.695\, {\cal D}_1 + 3784.97\, {\cal D}_0 + 1041.041\, L_1^4
  \\ & & \mbox{} \!\!\!
    - 8021.15\, L_1^2 - 15556.5\, (2-x) - 16445.21\, L_0 
    + 1084.36\, L_0^3 
  \quad\quad  \nonumber
\eea
and 
\bea
\label{eq311}
  c_{3,3}^{A}(x,5) &\, =\, & c_{S,3}^{}(x,5)
    - 6560.902\, {\cal D}_1 - 2412.54\, {\cal D}_0 + 98.499\, L_1^3
  \nonumber \\ & & \mbox{} \!\!\!
    - 27899.69\, L_1 - 82015.71\, x^2 - 2983.82 - 61.43\, L_0^3
  \nonumber \\[1mm]
  c_{3,3}^{B}(x,5) &\, =\, & c_{S,3}^{}(x,5)
    - 4637.854\, {\cal D}_1 + 4317.29\, {\cal D}_0 + 1070.036\, L_1^4
  \\ & & \mbox{} \!\!\!
    - 8767.685\, L_1^2 - 15676.57 - 6543.88\, L_0
    + 276.32\, L_0^3 \:\: .
  \quad\quad  \nonumber
\eea
In all cases the average $1/2\: (c_{a,3}^{A} + c_{a,3}^{B})$ represents 
our central result.

We conclude this section by returning to the coefficients $B_2$ and 
$D_2$ entering the NNLL soft-gluon resummation of the quark coefficient
functions (\ref{eq29}) and structure-function evolution kernels 
(\ref{eq214}). If only one of these constants were present, say $D_2$, 
then this constant would be fixed by the consistency of Eq.~(\ref{eq29})
with the soft-gluon part $c_{S,2}^{}$ of the NNLO coefficient functions 
of ref.~\cite{ZvN}, more precisely by the coefficient $A_2^{(0)}$ in 
Eq.~(\ref{eq31}). 
Digressing for a moment, we note that this situation is actually 
realized for the (very closely related) NNLL soft-gluon resummations of 
the quark-antiquark annihilation contribution to the Drell-Yan cross 
section \cite{av00} and of the cross section for Higgs production via 
gluon-gluon fusion in the heavy top-quark limit. For these two 
processes the NNLO soft- and virtual-gluon contribution have been 
computed in refs.~\cite{DYsoft} and \cite{Hsoft}, respectively. 
In the present DIS case, however, this consistency conditions only 
implies the constraint (\ref{eq215}). An exact result for the 
coefficient $A_2^{(3)}$ of the three-loop coefficient functions 
(\ref{eq31}) would suffice to determine $B_2$ and $D_2$, as this 
coefficient is related to the combination $B_2 + 2D_2$ independent from 
Eq.~(\ref{eq215}). 

As discussed in the paragraph below that of Eq.~(\ref{eq35}), all our
approximations (shown for $N_f = 4$ in Fig.~2) include $A_2^{(3)}$. For
about 95\% of these approximations this coefficient falls into the 
range
\beq
\label{eq311a}
  A_2^{(3)} \:\simeq\: \left\{ \begin{array}{c} 
    -6800 \:\ldots\: -5800 \\[1mm] -6350 \:\ldots\: -5500 \\[1mm] 
    -5950 \:\ldots\: -5200 \end{array} \right\}
 \quad \mbox{ for } \quad
 N_f \: = \:\left\{ \begin{array}{c} 
     3 \\[1mm] 4 \\[1mm] 5 \end{array} \right\} \:\: .
\eeq
The comparison of these results to the expansion of Eq.~(\ref{eq29})
(using $g_3(\lam)$ of ref.~\cite{av00}) leads to the rather weak
constraints
\beq
\label{eq311b}
  B_2  \:\simeq\: \left\{ \begin{array}{c} 
    32 \:\ldots\: 87 \\[1mm] 42 \:\ldots\: 93 \\[1mm] 
    49 \:\ldots\: 98 \end{array} \right\}
 \quad \mbox{ for } \quad
 N_f \: = \:\left\{ \begin{array}{c} 
     3 \\[1mm] 4 \\[1mm] 5 \end{array} \right\} 
\eeq
which can to sufficient accuracy be combined to the estimate
\beq
\label{eq312}
  B_2 \:\simeq\: - P_1^{\,\delta} + \frac{1}{3}\, \xi \beta_0 
  P_0^{\,\delta} 
  \quad \mbox{ with } \quad
  8 \:\lsim\: \xi \:\lsim\: 12 \:\: .
\eeq
Here $P_{l-1}^{\,\delta}$ represent the coefficients of $\delta 
(1\! -\! x)$ in the $l$-loop quark splitting functions. Retaining the 
colour factors $C_A = N_c = 3$, $C_F = (N_c^2-1)/(2N_c)$ these 
coefficients read
\bea
\label{eq313}
  P_0^{\,\delta} &=& 3\, C_F \nonumber \\[2mm] 
  P_1^{\,\delta} &=& 
    C_F^2 \left( \frac{3}{2} + 24\,\zeta_3 - 12\,\zeta_2 \right)
  + C_F C_A \left( \frac{17}{6} - 12\,\zeta_3 + \frac{44}{3}\,\zeta_2 
            \right) \\
  & & \mbox{}
  - C_F N_f \left( \frac{1}{3} + \frac{8}{3}\,\zeta_2 \right)
  \nonumber
\eea
for our normalization (\ref{eq22a}) of the expansion parameter. 
%
%
%
\section{Convolutions for the N{\boldmath$^3$}LO evolution kernels}
%
\setcounter{equation}{0}
Besides the $(l\! +\! 1)$-loop splitting functions and the $l$-loop 
coefficient functions, the N$^l$LO evolution kernels (\ref{eq25}) 
involve simple and multiple convolutions of the coefficient functions
of lower order.
Required at N$^3$LO are the simple and double convolutions
of the one-loop coefficient functions $c_{a,1}^{}$ with themselves, 
$c_{a,1}^{\otimes 2}$ and $c_{a,1}^{\otimes 3}$, and the convolutions 
of the one- with the two-loop coefficient functions, $c_{a,1}^{}\!
\otimes\! c_{a,2}^{}$, see Eq.~(\ref{eq26}). Especially the latter lead 
to rather complex exact expressions.
These terms, however, do not require any attention if the evolution is 
carried out using the moment-space technique \cite{cmom}, as in 
\mbox{$N$-space} the convolutions reduce to products. On the other 
hand, many analyses of data on structure functions are performed using 
`brute-force' $x$-space programs for solving the evolution equations. 
For application in such programs we provide compact and accurate 
parametrizations of the convolution contributions to the evolution 
kernels up to N$^3$LO.

These approximations are derived analogously to those of the two-loop 
coefficient functions in ref.~\cite{NV1}: The +-distribution parts
are treated exactly (truncating irrational coefficients), see the 
appendix. The integrable $x\! <\! 1$ terms are fitted to the exact 
results for $ x \geq 10^{-6} $. Finally the coefficients of
$\delta (1-x)$ are slightly adjusted from their exact values using the
lowest integer moments. The resulting parametrizations deviate from the
exact results by no more than a few permille. This accuracy applies 
directly to Eqs.~(\ref{eq41})--(\ref{eq46}) as well as to their 
convolutions with typical hadronic input distributions. 

Using the abbreviations (\ref{eq32}) the simple convolutions of the 
one-loop coefficient functions for $F_{2,\rm NS}$ and $F_3$ can be 
written as 
\bea
\label{eq41}
   c_{2,1}^{\otimes 2} (x) &\! =\! & 
       256/9\: \DD_3 - 64\,\DD_2 - 283.157\,\DD_1 + 304.751\,\DD_0 
       + 346.213\, \delta (1-x)
       \nonumber \\[0.5mm]
   & & \mbox{}- 26.51\, L_1^3 + 192.9\, L_1^2 + 198.2\, L_1
       + 113.0\, L_1^2 L_0 
       \nonumber \\[0.5mm]
   & & \mbox{}- 1.230\, L_0^3 + 9.466\, L_0^2 + 32.45\, L_0 - 483.3\, x 
       - 410.5 
\eea
and 
\bea
\label{eq42}
   c_{3,1}^{\otimes 2} (x)
   &=& 256/9\: \DD_3 - 64\,\DD_2 - 283.157\,\DD_1 + 304.751\,\DD_0 
       + 345.993\, \delta (1-x) \nonumber \\[0.5mm]
   & & \mbox{}- 27.09\, L_1^3 + 162.1\, L_1^2 + 248.0\, L_1
       + 91.79\, L_1^2 L_0 
       \nonumber \\[0.5mm]
   & & \mbox{}- 1.198\, L_0^3 + 3.054\, L_0^2 + 65.54\, L_0 - 305.3\, x 
       - 335.7 \:\: .
\eea
The corresponding parametrizations for the double convolutions read
\bea
\label{eq43}
   c_{2,1}^{\otimes 3}(x) &\! =\! & 
      1024/9\:\DD_5 - 1280/3\:\DD_4 - 2757.883\,\DD_3 + 9900.585\,\DD_2
   \nonumber\\[0.5mm] & & \mbox{} 
       + 3917.516\,\DD_1 - 12573.13\,\DD_0 - 2851.0\, \delta (1-x)
       \\
   & & \mbox{} - 151.4\, L_1^5 + 118.9\, L_1^4 - 6155\, L_1^3 
       - 47990\, L_1^2 - 30080\, L_1
       + 6423\, L_1^2 L_0 
       \nonumber \\[0.5mm]
   & & \mbox{} - 0.35\, L_0^5 - 4.30\, L_0^4 - 106.7\, L_0^3 
       - 1257\, L_0^2 - 4345\, L_0 + 3618\, x + 8547 \quad
       \nonumber
\eea
and 
\bea
\label{eq44}
   c_{3,1}^{\otimes 3}(x) &\! =\! & 
      1024/9\:\DD_5 - 1280/3\:\DD_4 - 2757.883\,\DD_3 + 9900.585\,\DD_2
   \nonumber\\[0.5mm] & & \mbox{} 
       + 3917.516\,\DD_1 - 12573.13\,\DD_0 - 2888.1\, \delta (1-x)
       \\
   & & \mbox{} - 138.4\, L_1^5 + 409.0\, L_1^4 - 1479\, L_1^3 
       - 24700\, L_1^2 + 9646\, L_1
       -10080\, L_1^2 L_0 
       \nonumber \\[0.5mm]
   & & \mbox{} - 0.119\, L_0^5 + 3.126\, L_0^4 + 84.84\, L_0^3 
       + 288.7\, L_0^2 + 264.87\, L_0 + 15410\, x + 15890 \: . 
       \nonumber
\eea

The convolutions $c_{a,2}^{}\otimes c_{a,1}^{}$ for $F_{2,\rm NS}$
in electromagnetic DIS and for the charged-current combination
$F_{3}^{\,\nu + \bar{\nu}}$ are parametrized as
\bea
\label{eq45}
 \lefteqn{ [ c_{2,2}^{}\otimes c_{2,1}^{} ](x) = } \nonumber \\[1mm]
   & & 1536/27\:\DD_5 - 343.702\,\DD_4 - 633.29\,\DD_3 
       + 5958.86\,\DD_2 - 6805.10\,\DD_1 - 2464.47\,\DD_0
       \nonumber \\[0.5mm]
   & & \mbox{} - 101.7\, L_1^5 - 155.1\, L_1^4 - 6553\, L_1^3 
       - 23590\, L_1^2 - 10620\, L_1
       + 9290\, L_1^2 L_0 - 0.35\, L_0^5
       \nonumber \\[0.5mm]
   & & \mbox{} + 0.64\, L_0^4 + 92.93\, L_0^3 + 761.9\, L_0^2 
       + 2450\, L_0 - 1251\, x + 6286 + 8609.2\, \delta (1-x)
       \nonumber \\[1.5mm]
   & & \hspace*{-2mm}\mbox{} + N_f\: \Big\{ 
       7.0912\,\DD_4 - 55.3087\,\DD_3 + 18.629\,\DD_2 
           + 619.865\,\DD_1 - 584.260\,\DD_0 
       \nonumber \\[0.5mm]
   & & \mbox{} - 11.71\, L_1^4 + 60.82\, L_1^3 - 618.0\, L_1^2 
       - 1979\, L_1 - 919.6\, L_1^2 L_0 + 0.48\, L_0^4
       \\
   & & \mbox{} - 1.08\, L_0^3 - 43.83\, L_0^2 
       - 125.5\, L_0 - 295.1\, x + 522.4\,
       - 809.14\, \delta (1-x) \Big\} \nonumber
\eea
and 
\bea
\label{eq46}
 \lefteqn{ [ c_{3,2}^{}\otimes c_{3,1}^{} ](x) = } \nonumber \\[1mm]
   & & 1536/27\:\DD_5 - 343.702\,\DD_4 - 633.29\,\DD_3
       + 5958.86\,\DD_2 - 6805.10\,\DD_1 - 2464.47\,\DD_0
       \nonumber \\[0.5mm]
   & & \mbox{} - 77.39\, L_1^5 + 289.1\, L_1^4 - 2823\, L_1^3
       - 12500\, L_1^2 + 25420\, L_1
       + 9515\, L_1^2 L_0 - 0.524\, L_0^5
       \nonumber \\[0.5mm]
   & & \mbox{} - 6.104\, L_0^4 + 39.23\, L_0^3 + 553.5\, L_0^2
       + 1393\, L_0 + 20080\, x + 2548 + 8478.2\, \delta (1-x)
       \nonumber \\[1.5mm]
   & & \hspace*{-2mm}\mbox{} + N_f\: \Big\{
       7.0912\,\DD_4 - 55.3087\,\DD_3 + 18.629\,\DD_2
           + 619.865\,\DD_1 - 584.260\,\DD_0
       \nonumber \\[0.5mm]
   & & \mbox{} - 14.30\, L_1^4 + 10.47\, L_1^3 - 775.2\, L_1^2
       - 2458\, L_1 - 392.9\, L_1^2 L_0 + 0.482\, L_0^4
       \\
   & & \mbox{} + 2.541\, L_0^3 - 41.04\, L_0^2
       - 223.9\, L_0 - 891.9\, x + 468.0\,
       - 803.43\, \delta (1-x) \Big\} \:\: . \nonumber
\eea
The corresponding results for the charged-current quantities 
$F_{2,\rm NS}$ and $F_{3}^{\,\nu - \bar{\nu}}$ are obtained by 
replacing the second and third line of Eqs.~(\ref{eq45}) and
(\ref{eq46}), respectively, by
\bea
  & &  \mbox{} - 109.2\, L_1^5 - 243.4\, L_1^4 - 6890\, L_1^3
       - 24000\, L_1^2 + 10840\, L_1 + 9144\, L_1^2 L_0
       - 0.45\, L_0^5
       \nonumber \\
  & &  \mbox{} + 1.80\, L_0^4 + 114.0\, L_0^3
       + 856.6\, L_0^2 + 2602\, L_0 - 711.6\, x + 6298
       + 8569.2\, \delta (1-x)
       \nonumber
\eea
and
\bea
  & &  \mbox{} - 77.39\, L_1^5 + 295.5\, L_1^4 - 2587\, L_1^3
       - 10580\, L_1^2 + 30580\, L_1 + 6461\, L_1^2 L_0
       - 0.404\, L_0^5
       \nonumber \\
  & &  \mbox{} - 5.525\, L_0^4 + 23.80\, L_0^3
       + 484.7\, L_0^2 + 1577\, L_0 + 22220\, x + 3349
       + 8485.0\, \delta (1-x) \:\: .
       \nonumber
\eea
%
%
%
\section{Numerical results for the scaling violations}
%
\setcounter{equation}{0}
In this section we illustrate the effect of the next-to-next-to-next-%
to-leading order (N$^3$LO) contributions to the physical evolution
kernels (\ref{eq25})--(\ref{eq27}) for the electromagnetic structure 
function $F_{2,\rm NS}$ and the charged-current combination $F_3^{\,\nu 
+\bar{\nu}}$ henceforth simply denoted $F_3$. Specifically, we will
discuss the logarithmic derivatives $\dot{F}_a \equiv d\ln F_a / 
d\ln Q^2$ calculated at a fixed reference scale $Q^2 = Q_{0\,}^2$ for 
the initial conditions
\beq
\label{eq51}
  F_{2,\rm NS}^{}(x, Q_0^2) \: =\: xF_{3}^{}(x, Q_0^2) \: =\: 
  x^{0.5} (1-x)^3 \:\: .
\eeq
The simple model shape (\ref{eq51}) incorporates the most important 
features of non-singlet $x$-distributions of nucleons. Its overall 
normalization is irrelevant for the logarithmic derivatives considered 
here. The reference scale $Q_{0\,}^2$ is specified via 
\beq
\label{eq52}
  \alpha_s (\mu_r^2\! = \! Q_0^2) \: = \: 0.2 
\eeq
irrespective of the order of the expansion. Eq.~(\ref{eq52}) 
corresponds to $Q_0^2 \simeq 30$ GeV$^2$, a scale typical for 
fixed-target DIS, for $\alpha_s (M_Z^2) \simeq 0.116$ beyond leading 
order (LO). The same input (\ref{eq51}) and (\ref{eq52}) is chosen at 
all orders and for both structure functions in order to facilitate a 
direct comparison of the various contributions to the evolution kernels
(\ref{eq25}). All graphical illustrations below refer to $N_f=4$ 
effectively massless quark flavours.

Before turning to the numerical results we have to specify our 
treatment of the four-loop splitting functions $P_{a,3}$. These 
functions enter Eq.~(\ref{eq25}) at order $\alpha_s^4$ (N$^3$LO) 
together with the three-loop coefficient functions (\ref{eq36}) --
(\ref{eq311}) and the convolutions (\ref{eq43}) -- (\ref{eq46}). As 
already mentioned in the introduction, the size of the two- and 
three-loop terms in the expansion of the non-singlet splitting 
functions strongly indicates that the effects of $P_{a,3}$ are very 
small in the $x$-region addressed by the present study, $x\!  >\! 
10^{-2}$. Hence a rather rough estimate of these quantities is 
sufficient here. We have checked that the [0/1] Pade approximation%
\footnote{$\,$A brief discussion of the Pad\'e summations and the 
 resulting higher-order approximations can be found in the next 
 section.} 
gives a reasonable, though not particularly accurate estimate of the 
three-loop non-singlet splitting functions $P_{a,2}^N$ in $N$-space for 
$N_f = 3\ldots 5$. Thus we choose the Mellin inverse of 
\beq
\label{eq53}
  P_{a,3}^N \:\,\simeq\: \eta\, 
  [ P_{a,3}^N ]^{}_{\, [1/1]\:\rm Pad\acute{e}} 
  \:\: , \quad \eta \: =\: 0\,\ldots\, 2
\eeq
as our estimate of $P_{a,3}(x)$, i.e., we assign a 100\% error to the
predictions of the [1/1] Pad\'e summation (the [0/2] Pad\'e results are 
similar). The results obtained by combining Eq.~(\ref{eq53}) for 
$\eta = 2$ with $c_{a,3}^{A}$ in Eqs.~(\ref{eq36}) -- (\ref{eq311}) are
denoted by N$^3$LO$_A$ in the figures below, those using $\eta = 0$
and $c_{a,3}^{B}$ by N$^3$LO$_B$. As in section 3 the central 
predictions $1/2\:({\rm N^3LO}_A+ {\rm N^3LO}_B)$ are not shown
separately.

\begin{figure}[p]
\vspace*{1mm}
\centerline{\epsfig{file=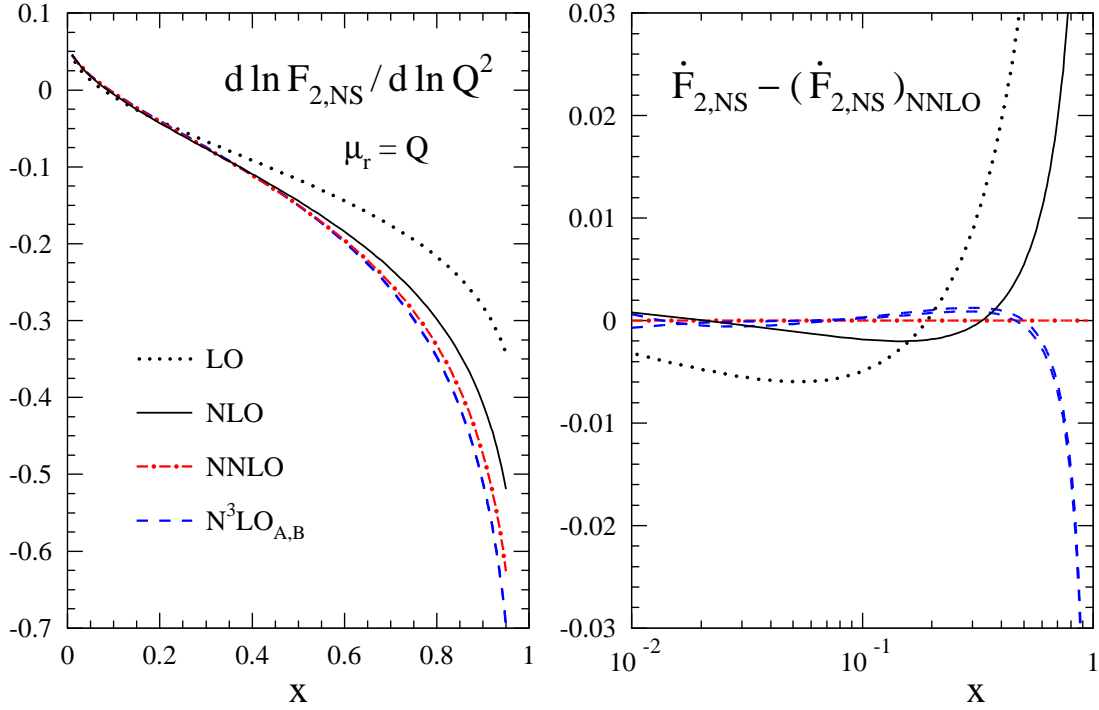,width=15cm,angle=0}}
\vspace{-1.5mm}
\caption{The perturbative expansion of the scale derivative 
 $\dot{F}_{2,\rm NS} \equiv d\ln F_{2,\rm NS\,} / d\ln Q^2$ of the 
 electromagnetic structure function $F_{2,\rm NS}$ at $\mu_r^2 = Q^2 
 \simeq 30 \mbox{ GeV}^2$ for the initial conditions specified in 
 Eqs.~(\ref{eq51}) and (\ref{eq52}). The differences between the 
 predictions at different orders in $\alpha_s$ are shown on a larger 
 scale in the right part.}
\vspace{1mm}
\end{figure}
\begin{figure}[p]
\vspace*{1mm}
\centerline{\epsfig{file=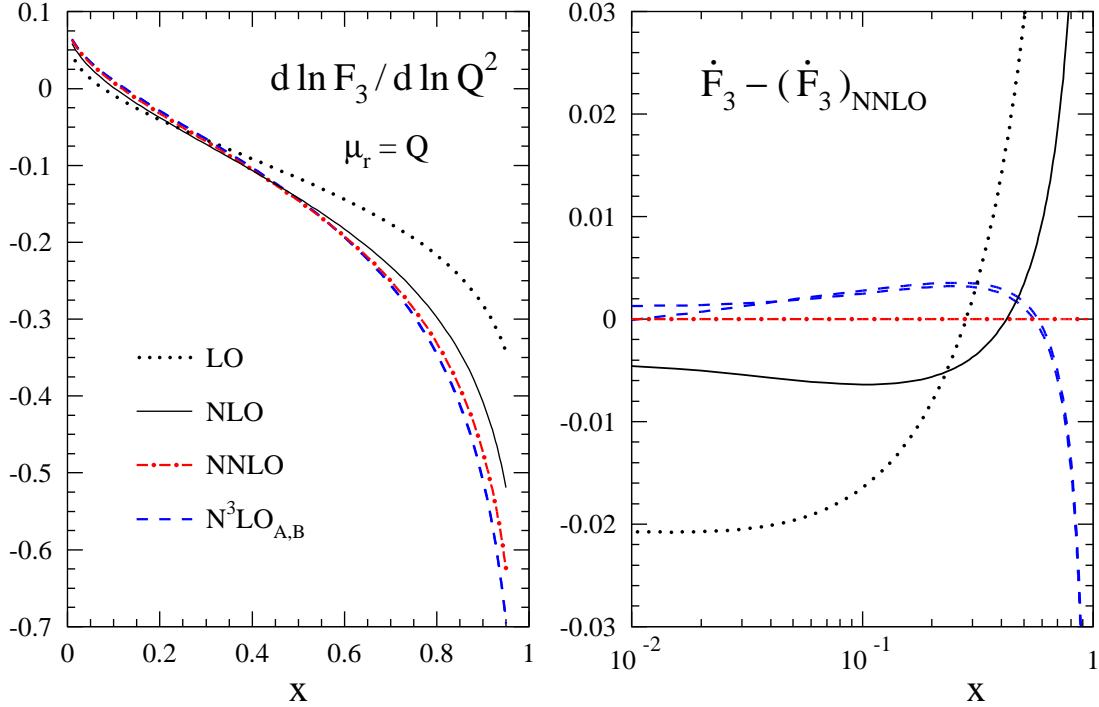,width=15cm,angle=0}}
\vspace{-1.5mm}
\caption{As Fig.~4, but for the  charged-current combination $F_3 
 \equiv F_3^{\,\nu + \bar{\nu}}$. In all figures the subscripts $A$ and 
 $B$ at N$^3$LO refer to the approximations discussed below 
 Eq.~(\ref{eq53}).}
\vspace{1mm}
\end{figure}

The logarithmic scale derivatives $\dot{F}_{2,\rm NS}$ and $\dot{F}_3$ 
resulting from Eqs.~(\ref{eq51}), (\ref{eq52}) are shown in the left 
parts of Fig.~4 and Fig.~5, respectively, for the standard choice 
$\mu_r^2 = Q^2$ of the renormalization scale. At $x\,\lsim\: 0.5$ the 
size of the NNLO and N$^3$LO corrections can barely be read on the 
scale of these graphs, therefore the differences $\dot{F}_a - 
(\dot{F}_a)_{\,\rm NNLO}$ are displayed on a larger scale in the left 
parts of both figures. The difference of the N$^3$LO$_A$ and 
N$^3$LO$_B$ results is very small down to $x \simeq 10^{-2}$ even on 
this enlarged scale, demonstrating that our approximations (\ref{eq36}) 
-- (\ref{eq311}) and (\ref{eq53}) are completely sufficient in this 
region of $x$.
For both structure functions the N$^3$LO corrections are large only at 
very large~$x$, where the kernels are dominated by the universal
soft-gluon contributions. Towards smaller $x$ the N$^3$LO effects
rapidly decrease, e.g., from 6\% at $x\! =\! 0.85$ to 2\% at 
$x\! =\! 0.65$. 
The corresponding NNLO contributions amount to about 12\% and 6\%, 
respectively. At $ 10^{-2}\,\lsim\, x \,\lsim\, 0.6$ the N$^3$LO 
corrections are particularly small for $F_{2,\rm NS}$.

\begin{figure}[tbh]
\vspace*{2mm}
\centerline{\epsfig{file=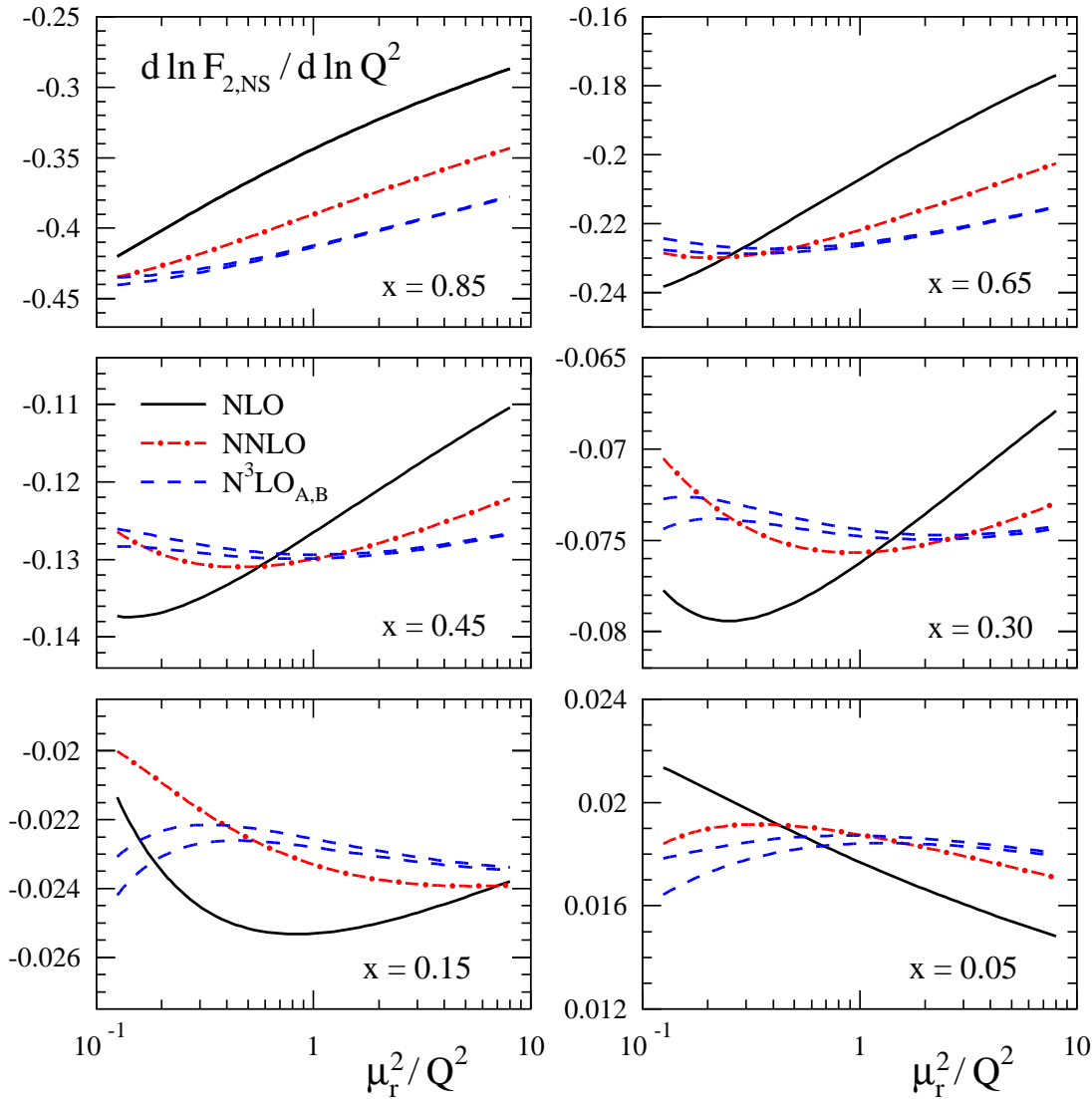,width=14.5cm,angle=0}}
\vspace*{-1mm}
\caption{The dependence of the NLO, NNLO and N$^3$LO predictions for
 $d\ln F_{2,\rm NS}/d\ln Q^2$ at $Q^2 = Q_0^2 \simeq 30\mbox{ GeV}^2$ 
 on the renormalization scale $\mu_r$ for six typical values of $x$.}
\end{figure}

\begin{figure}[tbh]
\vspace*{2mm}
\centerline{\epsfig{file=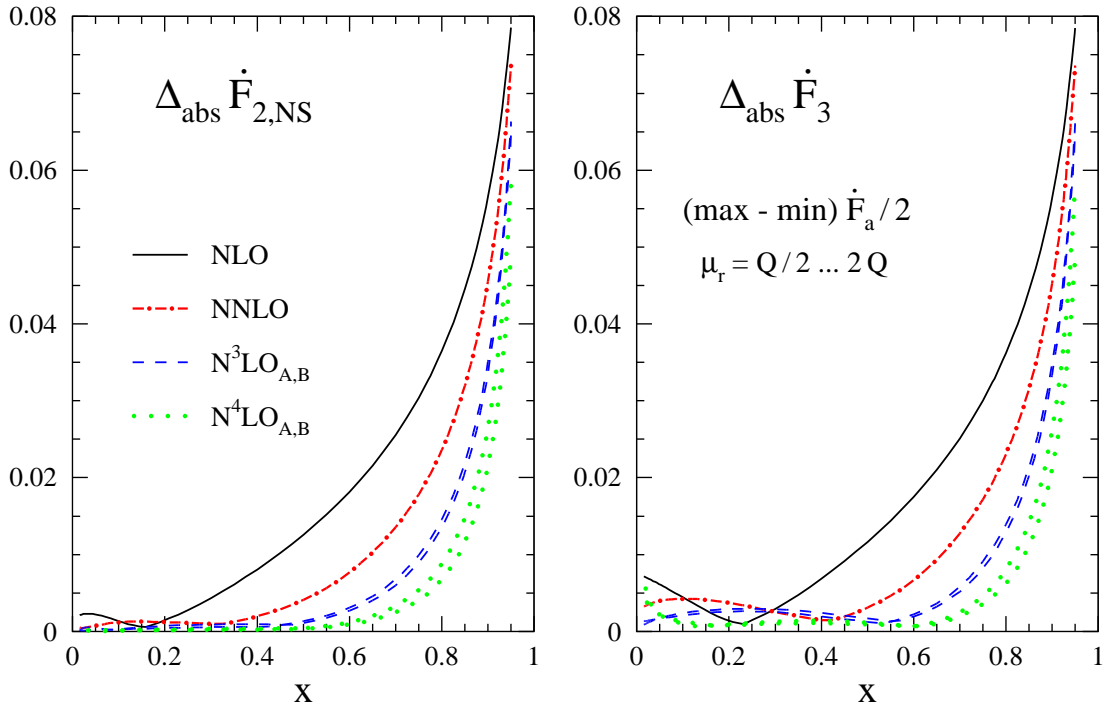,width=15cm,angle=0}}
\vspace*{-1mm}
\caption{The renormalization scale uncertainties, as estimated by the 
 quantities $\Delta_{\rm abs}\dot{F}_a$ defined in Eq.~(\ref{eq54}), of 
 the perturbative predictions for the scale derivatives of $F_{2,NS}$ 
 and $F_3$ displayed in Fig.~4 and Fig.~5, respectively. Also included 
 are the corresponding approximate N$^4$LO results derived in section 6 
 below.}
\end{figure}

The dependence of these scale derivatives on the renormalization 
scale\footnote{As already mentioned below Eq.~(\ref{eq22}), we use 
 the \MSb\ renormalization scheme throughout this study.}
is illustrated in Fig.~6 and Fig.~7. In the former figure the 
consequences of varying $\mu_r$ are shown for $\dot{F}_{2,\rm NS}$ at 
six representative values of $x$ (note that the scales of the ordinates
are different in all six parts). Here we vary $\mu_r$ over a rather 
wide range, $\frac{1}{8}\,Q^2\! \leq \mu_r^2 \leq 8 Q^2$, corresponding 
to $\, 0.29\,\gsim\,\alpha_s(\mu_r^2)\,\gsim\, 0.15\, $ for the initial 
condition (\ref{eq52}). In the latter figure we display the absolute 
scale uncertainties of $\dot{F}_{2,\rm NS}$ and $\dot{F}_3$ at 
$Q^2 = Q_0^2$, estimated by
\beq
\label{eq54}
 \Delta_{\rm abs} \dot{F}_{a} \, \equiv \, \frac{1}{2} 
 \left\{ \max \Big[ \dot{F}_{a}(\mu_r^2 = \frac{1}{4}\, Q^2 \ldots 
 4 Q^2) \Big] - \min \Big[ \dot{F}_{a}(\mu_r^2 = \frac{1}{4}\, Q^2 
 \ldots 4 Q^2) \Big] \right\} \:\: , 
\eeq
i.e., using the smaller conventional interval $\frac{1}{2}\, Q \ldots
2\, Q$ for $\mu_r$. Also shown here are the further improvements 
resulting from including the approximate $\alpha_s^5$ (N$^4$LO) 
contributions to Eqs.~(\ref{eq25}) -- (\ref{eq27}) discussed in the 
next section.

Our new N$^3$LO results represent a clear improvement over the NNLO 
stability \cite{NV1} for all $x$-values of Fig.~6 except for $x = 0.05$ 
(here, however, the absolute spread is very small, see Fig.~7), where 
the difference between the N$^3$LO$_A$ and N$^3$LO$_B$ results at small 
$\mu_r$ becomes comparable to the $\mu_r$-variation at NNLO. This 
enhanced sensitivity at small scales is due to the larger values of 
$\alpha_s$ up to almost 0.3, which enter the approximate contributions 
to Eq.~(\ref{eq25}) as $\alpha_s^4$. The present approximation 
uncertainties of the N$^3$LO results are actually dominated by the 
estimate (\ref{eq53}) for the four-loop splitting functions, not by the 
residual uncertainties of the three-loop coefficient functions 
quantified in Eqs.~(\ref{eq36})--(\ref{eq311}).
 
The results shown in Fig.~7 correspond to relative uncertainties 
$(\Delta_{\rm abs} \dot{F}_{a})/ \dot{F}_{a}$ of 8\% at NNLO and 5\% 
and 3\% at N$^3$LO and N$^4$LO, respectively, for both $F_{2,\rm NS}$ 
and $F_3$ at $x = 0.85$. The corresponding figures at $x = 0.65$ 
read 5\% (NNLO), 2\% (N$^3$LO) and 1\% (N$^4$LO). These scale 
uncertainties are rather similar to the relative size of the 
highest-order contributions at $\mu_r^2 = Q^2$ in Fig.~4 and Fig.~5 
(for the N$^4$LO contribution see Fig.~11 in section~6). Hence the 
$\mu_r$-variation (\ref{eq54}) and the size of the last contribution 
included in Eq.~(\ref{eq25}) yield consistent uncertainty estimates
in this region of $x$. At smaller $x$ the absolute scale uncertainties 
are very small at N$^3$LO and N$^4$LO. For $\Delta_{\rm abs} 
\dot{F}_{2,\rm NS}$ values even below 0.001 are reached for $x\leq 0.5$ 
at N$^3$LO and $x \leq 0.6$ at N$^4$LO. 

We conclude this section by illustrating the effect of the higher-order
terms in Eq.~(\ref{eq25}) on the determination of $\alpha_s$ from the
scaling violations of non-singlet structure functions. For this 
illustration we assume that $F_{2,\rm NS}$ and $F_3$ at $Q_0^2 \simeq 
30$ GeV$^2$ are given by Eq.~(\ref{eq51}) with negligible uncertainty. 
The resulting average N$^3$LO predictions of $\dot{F}_{2,\rm NS}$ and 
$\dot{F}_{3}$ for $\mu_r^2 = Q_{0\,}^2$ and $\alpha_s = 0.2$ are 
employed as model data at $x_k = 0.1\, k- 0.05$ with $k = 1,\ldots, 8$. 
Roughly following the experimental pattern, we assign errors of 0.005 
for $k = 2, \ldots\, 6$, of 0.01 for $k = 1, 7$ and of 0.02 for $k = 8$ 
to these data points (for Eq.~(\ref{eq55}) only the relative size of 
these errors is relevant). Again already including the N$^4$LO estimate 
obtained in the next section, the fits of $\alpha_s(Q_0^2)$ to these 
model data yield
\bea
\label{eq55}
 \alpha_s(Q_0^2)^{\,}_{\rm NLO}\:\: & = & 
 \: 0.2080
 {\begin{array}{l} \scriptstyle + \:\: 0.021\\[-1mm]
 \scriptstyle - \:\: 0.013\:\:\end{array}}
 \: , \quad 0.2035
 {\begin{array}{l} \scriptstyle + \:\: 0.019\\[-1mm]
 \scriptstyle  -\:\: 0.011\end{array}} 
 \nonumber\\[0.5mm]
 \alpha_s(Q_0^2)^{\,}_{\rm NNLO} 
 & = &
 \: 0.2010
 {\begin{array}{l} \scriptstyle + \:\: 0.008\\[-1mm]
 \scriptstyle  -\:\: 0.0025\end{array}}
 \: , \quad 0.1995
 {\begin{array}{l} \scriptstyle + \:\: 0.0065\\[-1mm]
 \scriptstyle  -\:\: 0.0015\end{array}}
 \nonumber\\[0.5mm]
 \alpha_s(Q_0^2)^{\,}_{\rm N^3LO}\: 
 & = &
 \: 0.2000
 {\begin{array}{l} \scriptstyle + \:\: 0.003 \\[-1mm]
 \scriptstyle  -\:\: 0.001\:\:\end{array}}
 \: , \quad 0.2000
 {\begin{array}{l} \scriptstyle + \:\: 0.0025 \\[-1mm]
 \scriptstyle  -\:\: 0.0005 \end{array}}
 \\[0.5mm]
 \alpha_s(Q_0^2)^{\,}_{\rm N^4LO}\:
 & = &
 \: 0.2000
 {\begin{array}{l} \scriptstyle + \:\: 0.0015 \\[-1mm]
 \scriptstyle  -\:\: 0.0005 \end{array}}
 \: , \quad 0.2005
 {\begin{array}{l} \scriptstyle + \:\: 0.0015 \\[-1mm]
 \scriptstyle  -\:\: 0.0005 \end{array}} \nonumber
\eea
where the first column refers to $F_{2,\rm NS}$ and the second to 
$F_3$. The central values represent the respective results for  
$\mu_r^2 = Q_0^2$, and the errors are due to the renormalization scale 
variation $\frac{1}{4}\, Q^2\leq \mu_r^2\leq 4\, Q^2$, for the N$^3$LO 
and N$^4$LO cases combined with the approximation uncertainties. 
Unlike the NNLO terms, the N$^3$LO and N$^4$LO corrections do not cause
significant shifts of the central values, but just lead to a reduction
of the $\mu_r$ uncertainties which reach about $\pm 1\%$ at N$^3$LO.
The difference of the NLO and NNLO central results for $F_3$ is half as
large as that for $F_{2,\rm NS}$. This effect is due to larger positive
corrections to the logarithmic derivative at $x < 0.4$ in the former 
case (see Figs.~4 and 5), which counteract the effect of the negative 
corrections at large $x$ in the fit. As far as Eq.~(\ref{eq55}) can be
compared to the fits of real data in refs.~\cite{KKPS,SaYn} (where 
higher-twist contributions affecting the central values are included), 
our finding are consistent with those results.
%
%
%
\section{Resummations and optimizations}
%
\setcounter{equation}{0}
Finally we address the predictions of the soft-gluon resummation, the
ECH and PMS scheme optimizations, and the Pad\'e approximations for the 
physical evolution kernel $K_a$ in Eq.~(\ref{eq25}). Being especially 
interested in the region of large-$x\, /\,$large-$N$, where the 
higher-order corrections are large but similar for $F_{2,\rm NS}$ and 
$F_3$, we will for brevity focus on the former, more accurately 
measured quantity. 
Also in this section the numerical results are given for $N_f= 4$ and 
the initial conditions (\ref{eq51}) and (\ref{eq52}). We will mainly 
consider the predictions of the above-mentioned approaches at fixed 
order in $\alpha_s$, and only at the end briefly turn to the 
all-order results for the soft-gluon exponentiation and the 
Pad\'e summations.

The N$^l$LO predictions of the soft-gluon resummation for the kernels 
(\ref{eq25}) are given by the terms $a_s^{l+1} {\cal K}_{{\rm res},l}$ 
in Eq.~(\ref{eq214}). Recall that the leading logarithmic (LL), next-%
to-leading logarithmic (NLL) and next-to-next-to-leading logarithmic 
(NNLL) contributions behave as $\ln^{l+1\!} N$, $\ln^{l\!}N$ and (at 
$l\!\geq\! 2$) $\ln^{l-1\!}N$, respectively. The terms in the $l$-loop
coefficient functions $c_{a,l}$ proportional to $\ln^{k\!}N$ with 
$k= l\! +\! 2,\ldots ,2l\,$ cancel in the combinations (\ref{eq26}) for 
$l\!\geq\! 2$. This implies that, from $l\! =\! 5$ onwards, actually 
none of the four leading $\ln^{k\!}N$ terms of $c_{a,l}$ presently 
fixed by the soft-gluon exponentiation (\ref{eq29}) contributes to the 
N$^l$LO kernels (\ref{eq25}). Consequently, we expect a pattern for the 
numerical soft-gluon approximations to the physical kernels which is 
rather different from that discussed in ref.~\cite{av99} for the \MSb\ 
coefficient functions.

The cumulative effect of soft-gluon terms at NNLO ($\alpha_s^3$) and 
N$^3$LO ($\alpha_s^4$) is compared to the (approximate) full results in 
Fig.~8. The two solid curves in the right plot refer to the N$^3$LO$_A$
and N$^3$LO$_B$ approximation discussed below Eq.~(\ref{eq53}), the two
NNLL results to $\xi\! =\! 8$ and $\xi\! =\! 12$ in Eq.~(\ref{eq312}). 
For the (undisplayed) NLO contribution the LL and NNLL predictions are 
considerably smaller and larger, respectively, than the full result, 
whereas the inclusion of also the $a^2_s N^0$ term arising from soft 
and virtual emissions leads to a reasonable approximation. Combined 
with this situation, the results of Fig.~8 indicate that the number of 
soft-gluon logarithms required for a realistic approximation at N$^l$LO 
systematically increases with the order $l$: The full NLO, NNLO, and 
N$^3$LO curves run between the LL and NLL, close to the NLL, and 
between the NLL and NNLL results, respectively.
The NNLL soft-gluon contribution may thus be expected to represent a 
reasonable estimate for the N$^4$LO ($\alpha_s^5)$ term of 
Eq.~(\ref{eq25}) at large $x\, /\,$large $N$. As shown in Fig.~9$\,$%
\footnote{Already Fig.~3 demonstrates that the universal soft-gluon 
 terms do not provide a good approximation for $x<0.7$, hence the 
 corresponding $x$-space results in Figs.~9 and 12 are shown only at 
 larger $x$.},
however, the spread due to the present uncertainty (\ref{eq312}) of the 
parameters $B_2$ and $D_2$ entering Eq.~(\ref{eq214}) is unfortunately 
rather large. Moreover, even if with this uncertainty removed, e.g., by 
a future exact calculation of the three-loop coefficient functions, 
Figs.~8 and 9 indicate that the soft-gluon resummation can hardly be 
expected to provide accurate information on the N$^4$LO term, even for
moments as large as $N \simeq 30$.  
 
\begin{figure}[p]
\vspace*{1mm}
\centerline{\epsfig{file=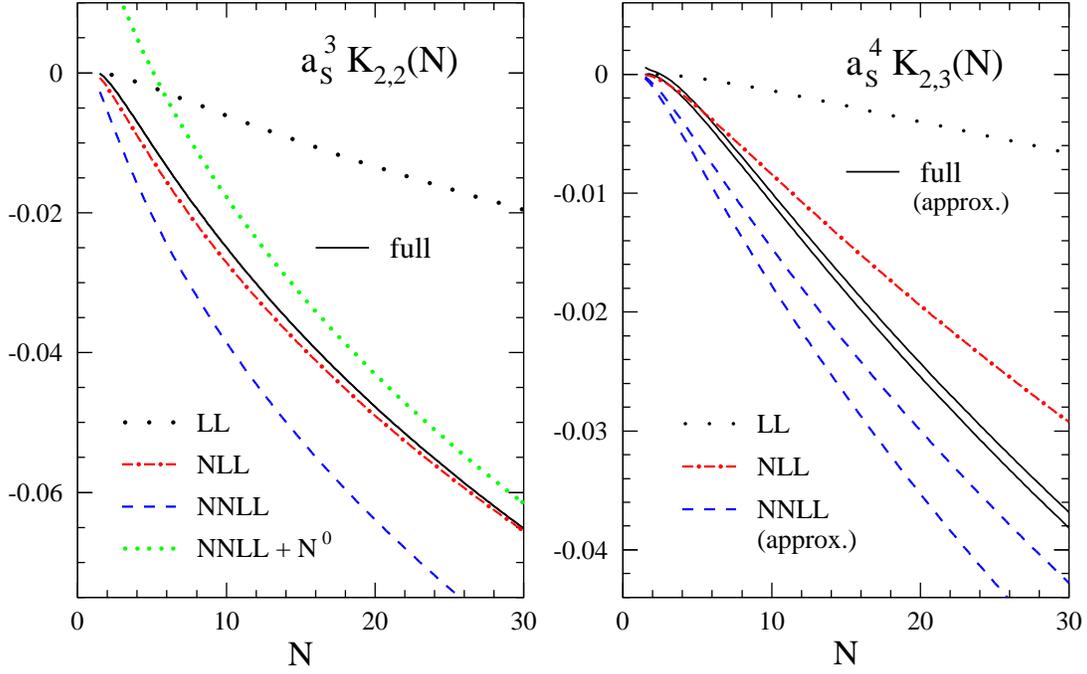,width=14.5cm,angle=0}}
\vspace{-1.5mm}
\caption{The successive soft-gluon approximations for the NNLO (left) 
 and N$^3$LO (right) contributions to the moments (\ref{eq28}) of the
 evolution kernel (\ref{eq25}) at $\mu_r^2 = Q^2$, compared with the 
 full results for $F_{2,\rm NS}$ addressed in the previous section. 
 Besides the $\ln^n N$ terms of $K_{\rm res,2}$ in  Eq.~(\ref{eq214}), 
 also the $N^0$ contribution is included for the NNLO case.}
\vspace{1mm}
\end{figure}
\begin{figure}[p]
\vspace*{1mm}
\centerline{\epsfig{file=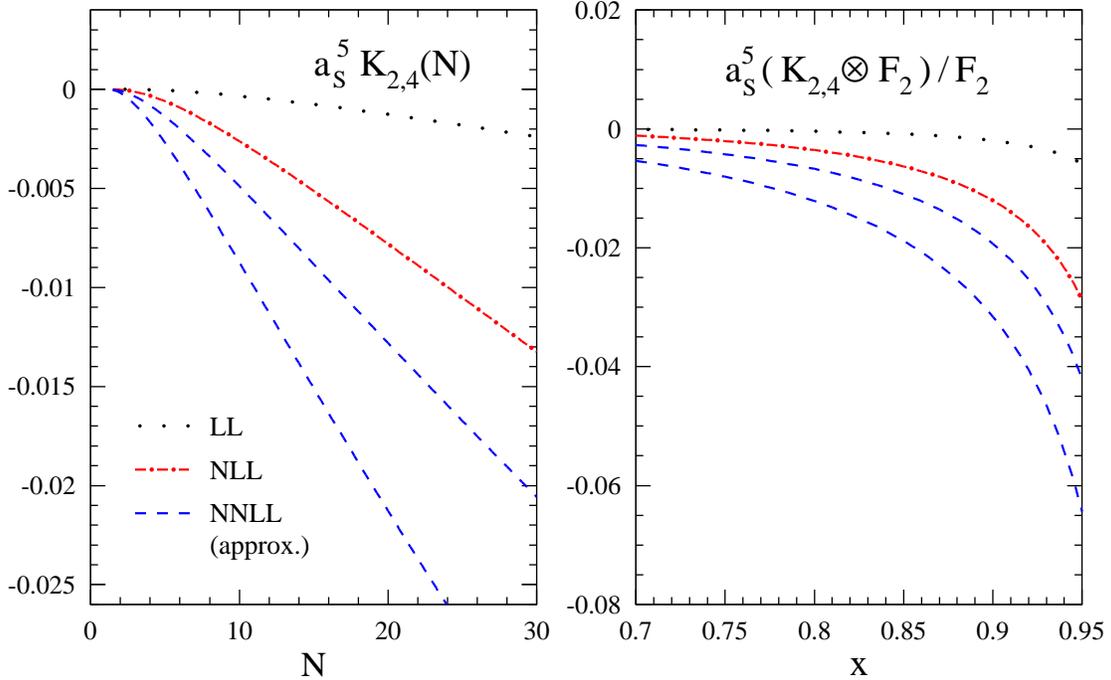,width=14.5cm,angle=0}}
\vspace{-1.5mm}
\caption{The results of the soft-gluon resummation (\ref{eq214}) for 
 the N$^4$LO contribution to the kernel (\ref{eq25}). The left part 
 corresponds to the right plot of Fig.~8, in the right part the 
 resulting large-$x$ predictions are shown for the $\alpha_s^5$ 
 corrections to the results of Fig.~4.}
\vspace{1mm}
\end{figure}

The renormalization-scheme optimizations assume that that the 
higher-order corrections to the N$^l$LO physical kernels 
${\cal K}_{a}^{(l)N}$ in $N$-space given by
\beq
\label{eq61}
  \frac{d \ln {\cal F}_{a}^{N}}{d\ln Q^2} \: = \:  {\cal K}_{a}^{(l)N}
  \: =\: a_s K_0^N \, ( 1 + a_s r_{a,1}^N + \ldots + a_s^l r_{a,l}^N )
\eeq
are small in a certain `optimal' scheme. The principle of minimal
sensitivity (PMS) proposed in ref.~\cite{PMS} selects this scheme by 
the requirement
\beq
\label{eq62}
  \frac{d^{\,}{\cal K}_{a}^N({\rm RS})}{d({\rm RS})} \: = \: 0 \:\: ,
\eeq
where $d/d({\rm RS})$ abbreviates the derivatives with respect to the
$l$ independent parameters specifying the renormalization scheme at 
N$^l$LO. In the effective charge (ECH) method of ref.~\cite{ECH}, on 
the other hand, these parameters are chosen such that
\beq
\label{eq63}
  r_{a,1}^N \: =\: \ldots \: =\:  r_{a,l}^N \: =\:  0 \:\: .
\eeq
Assuming that in these schemes the next terms $r_{a,l+1}^N$ are not 
just small but vanishing, the transformation back to \MSb\ (or any
other scheme) leads to the respective PMS and ECH predictions for this
quantity in terms of $r_{a,1}^N \:\ldots\: r_{a,l}^N$ and the 
coefficients (\ref{eq22}) of the $\beta$-function. Up to $r_4{}$ these 
predictions are explictly given in Eqs.~(6)~--~(11) and (13)~--~(17) of 
ref.\ \cite{KS95}, thus we refrain from repeating them here.

Another approach for estimating the higher-order corrections is 
provided by the Pad\'e summation of the perturbation series, for QCD 
in detail discussed, e.g., in refs.~\cite{Pade}. In this method 
${\cal K}_{a}^{(l)N}$ in Eq.~(\ref{eq61}) is replaced by
\beq
\label{eq64}
 {\cal K}^{N}_{\, a,\, [\cal{N}/\cal{D}]} \: = \: a_s K_0 \, 
 \frac{1 + a_s p_{a,1}^N + \ldots + a_s^{\cal N} p_{a,{\cal N}}^N}
 {1 + a_s q_{a,1}^N + \ldots + a_s^{\cal D} q_{a,{\cal D}}^N}
\eeq
with
\beq
\label{eq65}
 {\cal D} \:\geq\: 1 \quad \mbox{and} \quad 
 {\cal N} + {\cal D} \: = \: l \:\: .
\eeq
The determination of the parameters $p_i$ and $q_j$ from the $r_1 
\,\ldots\, r_l$ of Eq.~(\ref{eq61}) are automatized in programs for
symbolic manipulation such as {\sc Maple} \cite{Maple}. Expanding 
${\cal K}^{N}_{\, a,\, [\cal{N}/\cal{D}]}$ to order $l\! +\! 1$ then 
yields the $[\cal{N}/\cal{D}]$ Pad\'e predictions for the N$^{l+1}LO$ 
coefficients $r_{a,l+1}^N$. Also these predictions need not to be
written down here.
Beyond the second-order results 
%
%
there is no obvious relation between the predictions of the scheme
optimizations and those of the Pad\'e approximations. Consistent 
result of these methods for $r_{l>2}$ are thus usually considered as 
evidence of the approximate correctness of these predictions 
\cite{Pade}. 

\begin{figure}[p]
\vspace*{1mm}
\centerline{\epsfig{file=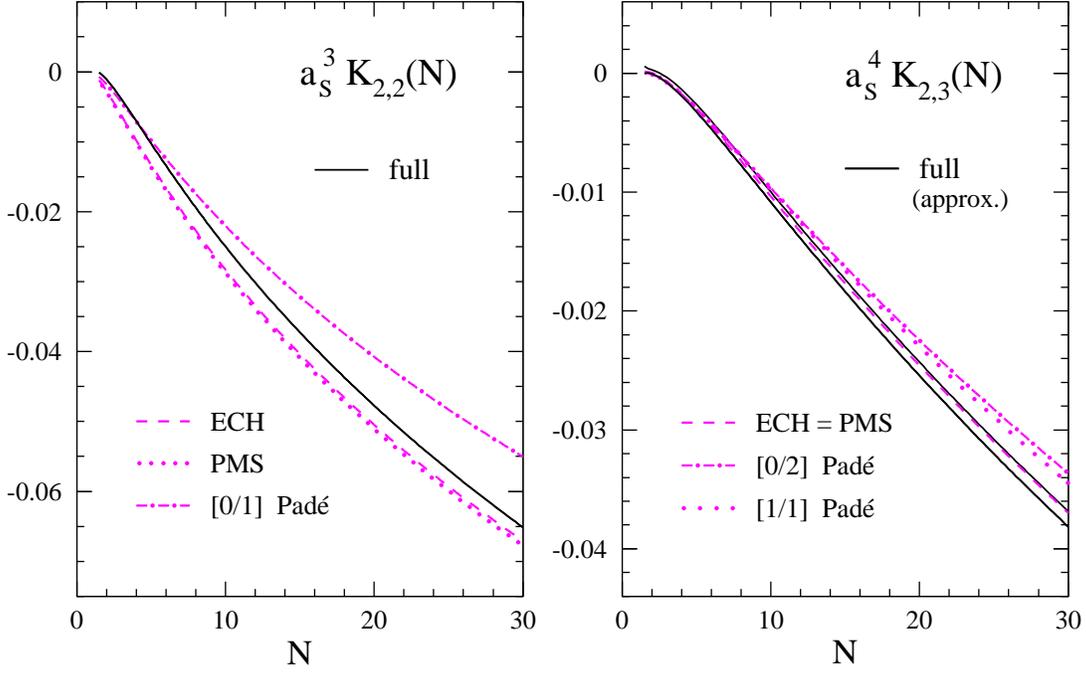,width=14.5cm,angle=0}}
\vspace{-1.5mm}
\caption{The PMS, ECH and Pad\'e estimates of the NNLO (left) and 
 N$^3$LO (right) parts of the $N$-space evolution kernel for 
 $F_{2,\rm NS}$ at $\mu_r^2 = Q^2$, compared with the full results 
 illustrated in $x$-space in section 5. The scales of the graphs are 
 the same as in Fig.~8.}
\vspace{1mm}
\end{figure}
\begin{figure}[p]
\vspace*{1mm}
\centerline{\epsfig{file=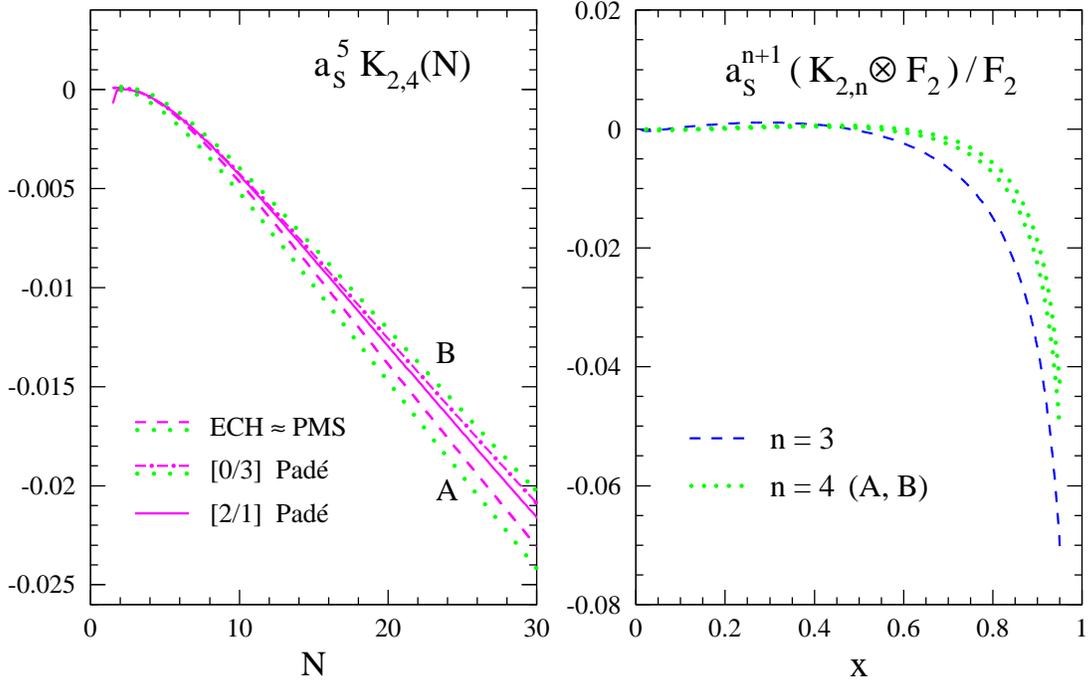,width=14.5cm,angle=0}}
\vspace{-1.5mm}
\caption{The PMS, ECH and Pad\'e estimates of the N$^4$LO contributions
 to the evolution kernel (\ref{eq25}) for $F_{2,\rm NS}$. The left part 
 is analogous to Fig.~10, in the right part the resulting $\alpha_s^5$ 
 corrections to the result in Fig.~4 are compared to the $\alpha_s^4$ 
 (N$^3$LO) contribution. The scales of the graphs are the same as in 
 Fig.~9.}
\vspace{1mm}
\end{figure}

The PMS, ECH and Pad\'e results for the NNLO and N$^3$LO $N$-space 
kernels (\ref{eq61}) are compared in Fig.~10 to the (approximate) full 
results already shown, on the same scale, in Fig.~8. Disregarding large 
relative, but small absolute deviations at NNLO for small~$N$, the PMS 
and ECH results (which are very similar at NNLO and identical at 
N$^3$LO) represent good approximations at both orders. The Pad\'e 
approximations are somewhat smaller, however, this offset seems to 
decrease with the order in $\alpha_s$.
In the left part of Fig.~11 we present the corresponding N$^4$LO
predictions. The inner three curves have been derived from the central
N$^3$LO results of section 5. The PMS and ECH are again very similar,
they are not shown separately. The impact of the present uncertainty of 
the N$^3$LO kernels, dominated by the estimate (\ref{eq53}) of the 
four-loop splitting functions, is included in the two dotted curves 
which represent our final estimate for the N$^4$LO term and its 
uncertainty. In the right part of Fig.~10 the N$^4$LO corrections to 
the results of Fig.~4 are compared to the N$^3$LO contribution. Within 
the large uncertainties of the latter, these results are consistent 
with the NNLL soft-gluon prediction shown in Fig.~9. 
The consequences of including the N$^4$LO estimates have been presented 
in Fig.~7 and Eq.~(\ref{eq55}), respectively, for the 
renormalization-scale stability and the determination of~$\alpha_s$.

Finally the $\alpha_s^{l>5}$ infinite-order predictions of the soft-%
gluon resummation (\ref{eq214}) (using the minimal prescription contour 
\cite{cont}) and the Pad\'e approximations (\ref{eq64}) are compared in 
Fig.~12. For the present uncertainties (\ref{eq312}) 
--- the curves in the figure refer to $\xi\! =\! 8$ (upper), $D_2\! =\! 
0$ in Eq.~(\ref{eq215}) (middle) and $\xi\! =\! 12$ (lower) ---
it is not possible to draw any conclusions from the soft-gluon result. 
The Pad\'e summation, on the other hand, provides rather definite 
predictions: The terms beyond $\alpha_s^{5}$ can be expected to have a 
very small impact at $x\,\lsim\, 0.75$. For our standard reference 
value $\alpha_s = 0.2$ their effect reaches about the size of the 
N$^4$LO and N$^3$LO contributions at $x\simeq 0.9$ and $x\simeq 0.95$, 
respectively.

\begin{figure}[hbt]
\vspace*{2mm}
\centerline{\epsfig{file=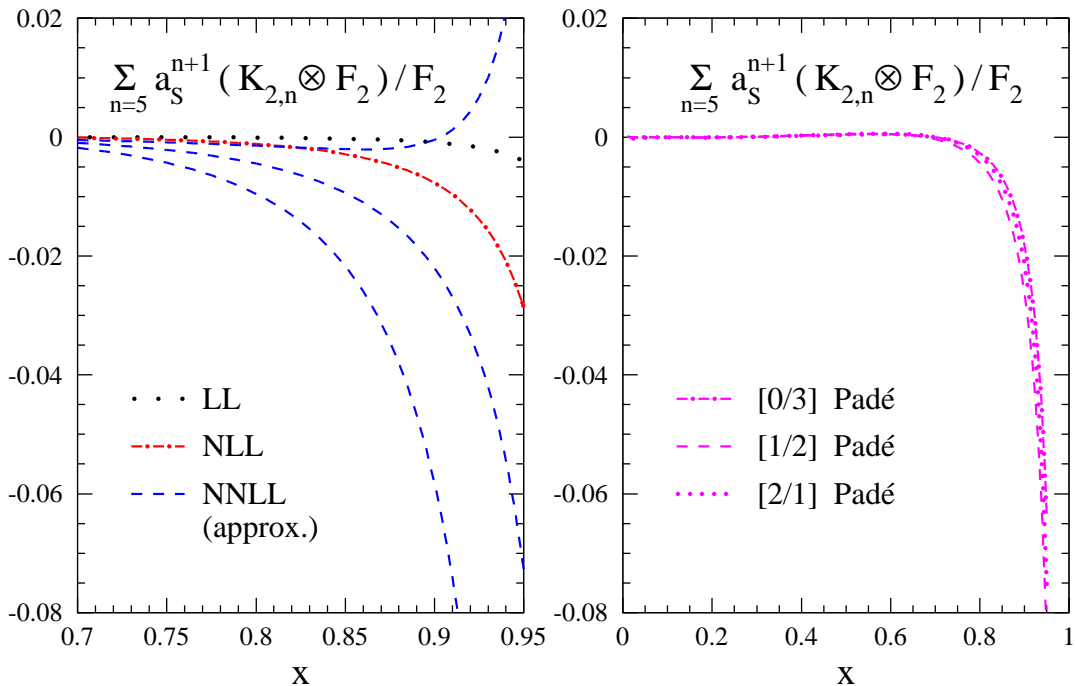,width=14.5cm,angle=0}}
\vspace{-1.5mm}
\caption{Predictions of the soft-gluon resummation (\ref{eq214}) and 
 the Pad\'e approximations (\ref{eq64}) for the contributions beyond
 N$^4$LO to the logarithmic scale derivative of $F_2 \equiv 
 F_{2,\rm NS}$.} 
\vspace{1mm}
\end{figure}
%
%
%
\section{Summary}
%
%
We have investigated the predictions of massless perturbative QCD for 
the scaling violations of the most important non-singlet structure
functions in unpolarized DIS, extending our previous NNLO results 
\cite{NV1} to N$^3$LO and N$^4$LO for the region $x\! >\! 10^{-2}$. 
The main objective of this extension is to reduce the theoretical 
uncertainty of determinations of $\alpha_s$ from inclusive DIS to about 
$1\%$, an accuracy which is sufficient to make full use of present and
future structure function measurements. 
Our results also facilitate improved determinations of power-suppressed
contributions to the structure-function evolution by fits to data,
especially at large $x$ where the uncertainties are still sizeable at 
NNLO. 

The new ingredients entering the N$^3$LO physical evolution kernels 
are the four-loop splitting functions and the three-loop coefficient 
functions. The impact of the former quantities is expected to be very 
small in the \MSb\ scheme at $x\! >\! 10^{-2}$; it has been estimated 
by a Pad\'e approximation assigned a 100\% uncertainty. 
For the latter quantities we have derived approximate expressions based 
on the available integer moment \cite{moms,mnew} and soft gluon 
\cite{sglue,av99} results. The effect of these functions is very well 
under control at $x\! >\! 10^{-2}$, and almost perfectly at $x \,\gsim 
\, 0.1$. In fact, the uncertainty of the splitting functions dominates 
the small residual uncertainties of the evolution kernels. Hence the 
accuracy of our present N$^3$LO results will be superseded only by a 
future four-loop calculation.

We have also studied the predictions of the NNLL soft-gluon resummation 
\cite{av00} and of the Pad\'e, PMS and ECH approximations \cite
{Pade,PMS,ECH}. Presently the predictions of the resummation for the 
physical evolution kernels beyond N$^3$LO (in any case applicable only 
at $x \gsim 0.8$) suffer from the incomplete determination of the 
soft-gluon parameters $B_2$ and $D_2$, a problem which will be removed
by forthcoming exact calculation of the three-loop coefficient 
functions \cite{MV2}. The Pad\'e, PMS and ECH approximations are found 
to agree rather well with the NNLO and N$^3$LO results for the 
evolution kernels; these approaches seem to provide reliable 
predictions of the effects at N$^4$LO and beyond.

For $\alpha_s \,\lsim\, 0.2$ the N$^3$LO and N$^4$LO corrections at 
$\mu_r \simeq Q$ are very small at $x\!<\! 0.6$ and $x\!<\! 0.8$, 
respectively, especially for the most accurately measured structure 
function $F_2$. Consequently the central values of $\alpha_s$ 
determined from the non-singlet scaling violations hardly change any 
more once the larger NNLO terms have been included, see also ref.\ 
\cite{KKPS}.
The scale uncertainty of the resulting $\alpha_s(M_Z^2)$ is reduced to 
the unproblematic level of less than 1\% at N$^3$LO and 0.5\% at 
N$^4$LO. In order to ensure an overall theoretical accuracy of about 
1\% also the heavy quark (especially charm) mass effects need to be 
controlled with this precision. We will address this point in a 
forthcoming publication. 

{\sc Fortran} subroutines of our approximations of the three-loop
coefficient functions in section 3 and of the parametrizations of the
convolutions entering the evolution kernels in section 4 can be found 
at {\tt http://www.lorentz.leidenuniv.nl/$\sim$avogt}. 
%
%
\section*{Acknowledgment}
%
%
We are grateful to J. Vermaseren for communicating the fourteenth 
moment of the three-loop coefficient function for $F_2$ to us prior to
publication.
This work has been supported by the European Community TMR research
network `Quantum Chromodynamics and the Deep Structure of Elementary 
Particles' under contract No.~FMRX--CT98--0194.
%
%
%
\section*{Appendix: Convolutions of +-distribution} 
%
\setcounter{equation}{0}
\renewcommand{\theequation}{A.\arabic{equation}}
The convolutions $\DD_i \otimes \DD_j$ of the +-distributions 
(\ref{eq32}) for $i+j \leq 4$ are given by
\bea
\label{eqa1}
  \DD_0 \otimes \DD_0 &=&
  2\, \DD_1 - \zeta_2\, \delta (1\! -\! x) 
  \nonumber\\[2mm] 
  \DD_1 \otimes \DD_0 &=&
  \frac{3}{2}\,\DD_2 - \zeta_2 \,\DD_0 
  + \zeta_3\, \delta (1\! -\! x) 
  \nonumber\\[2mm]
  \DD_1 \otimes \DD_1 &=&
  \DD_3 - 2 \zeta_2 \,\DD_1 + 2 \zeta_3 \,\DD_0
  - \frac{1}{10}\, \zeta_2^2\, \delta (1\! -\! x) 
  \nonumber\\[2mm]
  \DD_2 \otimes \DD_0 &=&
  \frac{4}{3}\, \DD_3 - 2 \zeta_2 \,\DD_1 + 2 \zeta_3 \,\DD_0 
  - \frac{4}{5}\, \zeta_2^2\, \delta (1\! -\! x) 
  \nonumber\\[2mm]
  \DD_2 \otimes \DD_1 &=&
  \frac{5}{6}\, \DD_4 - 3 \zeta_2 \,\DD_2 + 6 \zeta_3 \,\DD_1
  - \zeta_2^2 \,\DD_0
  + \Big( 4 \zeta_5 - 2 \zeta_2 \zeta_3 \Big) \,\delta (1\! -\! x) 
  \nonumber\\[2mm]
  \DD_2 \otimes \DD_2 &=&
  \frac{2}{3}\, \DD_5 - 4 \zeta_2 \,\DD_3 + 12 \zeta_3 \,\DD_2
  - 4\zeta_2^2 \,\DD_1 + \Big( 16\zeta_5 - 8 \zeta_2 \zeta_3 \Big)
  \,\DD_0 
  \\[1mm] & & \mbox{}
  + \left( 4 \zeta_3^2 - \frac{46}{35}\, \zeta_2^3 \right) 
  \,\delta (1\! -\! x) 
  \nonumber\\[2mm]
  \DD_3 \otimes \DD_0 &=& 
  \frac{5}{4}\, \DD_4 - 3 \zeta_2 \,\DD_2 + 6 \zeta_3 \,\DD_1 
  - \frac{12}{5}\, \zeta_2^2 \,\DD_0 
  + 6 \zeta_5\, \delta (1\! -\! x) 
  \nonumber\\[2mm]
  \DD_3 \otimes \DD_1 &=&
  \frac{3}{4}\, \DD_5 - 4 \zeta_2 \,\DD_3 + 12 \zeta_3 \,\DD_2
  -\frac{27}{5}\, \zeta_2^2 \,\DD_1 + \Big(18 \zeta_5 
  - 6 \zeta_2 \zeta_3 \Big) \,\DD_0 
  \nonumber \\[1mm] & & \mbox{} 
  + \left( 3\zeta_3^2 - \frac{36}{35}\, \zeta_2^3 \right) 
  \,\delta (1\! -\! x)
  \nonumber\\[2mm]
  \DD_4 \otimes \DD_0 &=&
  \frac{6}{5}\, \DD_5 - 4 \zeta_2 \,\DD_3 + 12 \zeta_3 \,\DD_2
  -\frac{48}{5}\, \zeta_2^2 \,\DD_1 + 24 \zeta_5 \,\DD_0
  -\frac{192}{35}\, \zeta_2^3\, \delta (1\! -\! x) \quad
  \nonumber
\eea
up to integrable contributions dealt with numerically in section 4. 
Here $\zeta_l$ stand for the Riemann $\zeta$-function, and $\zeta_4$ 
and $\zeta_6$ have been expressed in terms of $\zeta_2$ and $\zeta_3$,
respectively. 
A convenient method to derive (and to extend, if required) 
eqs.~(\ref{eqa1}) is by using the relation between $\DD_i$ and the 
harmonic sums discussed, for example, in ref.~\cite{JB98}.
%
%
%

%

\newpage
\begin{thebibliography}{99}
%
%
\bibitem{PDG}   D.E. Groom et al., Particle Data Group, Eur.\ Phys.\ J.
                {\bf C15} (2000) 1, and refs.~therein 
%
\bibitem{FP82}  W.\ Furmanski and R.\ Petronzio, Z. Phys.\ {\bf C11} 
                (1982) 293
%
\bibitem{KKPS}  A.L. Kataev, A. Kotikov, G. Parente and A.V. Sidorov,
                Phys.\ Lett.\ {\bf B417} (1998)~374;\\
                A.L. Kataev, G. Parente and A.V. Sidorov, Nucl.\ 
                Phys.\ {\bf B573} (2000) 405; preprint 
                {\tt hep-ph/0012014} (CERN-TH-2000-343)
%
\bibitem{SaYn}  J. Santiago and F.J. Yndurain, Nucl.\ Phys.\ {\bf B563} 
                (1999) 45; \\
                preprint {\tt hep-ph/0102312} (FTUAM 01-01)
%
\bibitem{ZvN}   E.B. Zijlstra and W.L. van Neerven, Phys.\ Lett.\ 
                {\bf B272} (1991) 127; ibid.\ {\bf B273} (1991) 476;
                ibid.\ {\bf B297} (1992) 377; Nucl.\ Phys.\ {\bf B383}
                (1992) 525
%
\bibitem{MV1}   S. Moch and J.A.M. Vermaseren, Nucl.\ Phys.\ {\bf B573}
                (2000) 853
%
\bibitem{moms}  S.A. Larin, T. van Ritbergen, and J.A.M. Vermaseren,
                Nucl.\ Phys.\ {\bf B427} (1994) 41; \\
                S.A. Larin, P. Nogueira, T. van Ritbergen, and J.A.M.
                Vermaseren, Nucl.\ Phys.\ {\bf B492} (1997) 338
%
\bibitem{mnew}  A. Retey and J.A.M. Vermaseren,  preprint 
                {\tt hep-ph/0007294} (NIKHEF-2000-018);\\
                J.A.M. Vermaseren, private communication
%
\bibitem{lNf}   J. A. Gracey, Phys.\ Lett.\ {\bf B322} (1994) 141; \\
                J.F. Bennett and J.A. Gracey, Nucl.\ Phys.\ {\bf B517}
                (1998) 241
%
\bibitem{lowx}  S. Catani and F. Hautmann, Nucl.\ Phys.\ {\bf B427}
                (1994) 475;\\
                J. Bl\"umlein and A. Vogt, Phys.\ Lett.\ {\bf B370}
                1996) 149;\\
                V.S. Fadin and L.N. Lipatov, Phys.\ Lett.\ {\bf B429}
                (1998) 127, and refs.~therein;\\
                M. Ciafaloni and G. Camici, Phys.\ Lett.\ {\bf B430}
                (1998) 349
%
\bibitem{NV1}   W.L. van Neerven and A. Vogt, Nucl.\ Phys.\ {\bf B568}
                (2000) 263 
%
\bibitem{NV2}   W.L. van Neerven and A. Vogt, Nucl.\ Phys.\ {\bf B588}
                (2000) 345
%
\bibitem{NV3}   W.L. van Neerven and A. Vogt, Phys.\ Lett.\ {\bf B490} 
                (2000) 111
%
\bibitem{c2DY}  R. Hamberg, W.L. van Neerven and T. Matsuura, Nucl.\
                Phys.\ {\bf B359} (1991) 343; \\
                W.L. van Neerven and E.B. Zijlstra, Nucl.\ Phys.\
                {\bf B382} (1992) 
%
\bibitem{sglue} G. Sterman, Nucl.\ Phys.\ {\bf B281} (1987) 310; \\
                L. Magnea, Nucl.\ Phys.\ {\bf B349} (1991) 703;\\
                S. Catani and L. Trentadue, Nucl.\ Phys.\ {\bf B327}
                (1989) 323; ibid.\ {\bf B353} (1991) 183;\\
                S. Catani, G. Marchesini and B.R. Webber, Nucl.\
                Phys.\ {\bf B349} (1991) 635
%
\bibitem{av99}   A. Vogt, Phys.\ Lett.\ {\bf B471} (1999) 97
%
\bibitem{av00}   A. Vogt, Phys.\ Lett.\ {\bf B497} (2001) 228
%
\bibitem{Pade}   M.A. Samuel, J. Ellis and M. Karliner, Phys.\ Rev.\
                 Lett.\ {\bf 74} (1995) 4380;\\
                 J. Ellis, E. Gardi, M. Karliner and M.A. Samuel,
                 Phys.~Lett.\ {\bf B366} (1996) 268;\\
                 Phys.\ Rev.\ {\bf D54} (1996) 6986;\\
                 S.~J.~Brodsky et al., Phys.\ Rev.\ {\bf D56} (1997) 
                 6980
%
\bibitem{PMS}    P.~M.~Stevenson, Phys.~Rev.\ {\bf D23} (1981) 2916
%
\bibitem{ECH}    G. Grunberg, Phys.~Rev.\ {\bf D29} (1984) 2315
%
\bibitem{beta2}  O.V. Tarasov, A.A. Vladimirov, and A.Yu.\ Zharkov,
                 Phys.\ Lett.\ {\bf B93} (1980) 429; \\
                 S.A. Larin and J.A.M. Vermaseren, Phys.\ Lett.\
                 {\bf B303} (1993) 334
%
\bibitem{beta3}  T. van Ritbergen, J.A.M. Vermaseren and S.A. Larin, 
                 Phys.\ Lett.\ {\bf B400} (1997)~379
%
\bibitem{CFP}   G. Curci, W. Furmanski and R. Petronzio, Nucl.\ Phys.\
                {\bf B175} (1980) 27
%
\bibitem{spx1}   A. Gonzales-Arroyo, C. Lopez and F.J. Yndurain, Nucl.\
                 Phys.\ {\bf B153} (1979) 161;\\
                 G. P. Korchemsky, Mod.\ Phys.\ Lett.\ {\bf A4} (1989)
                 1257;\\
                 S. Albino and R.D. Ball, preprint {\tt hep-ph/0011133} 
                 (CERN-TH/2000-332)
%
\bibitem{DYsoft} T. Matsuura and W.L. van Neerven, Z. Phys.\ {\bf C38}
                 (1988) 623
%
\bibitem{Hsoft}  S.~Catani, D.~de Florian and M.~Grazzini,
                 preprint {\tt hep-ph/0102227} (CERN-TH-2001-044);\\
                 R.~V.~Harlander and W.~B.~Kilgore,
                 preprint {\tt hep-ph/0102241} (BNL-HET-01-6)
%
\bibitem{cont}   S. Catani, M.L. Mangano, P. Nason and L. Trentadue,
                 Nucl.\ Phys.\ {\bf B478} (1996) 273
%
\bibitem{hpol}   E. Remiddi and J.A.M. Vermaseren, Int.\ J.\ Mod.\ 
                 Phys.\ {\bf A15} (2000) 725
%
\bibitem{KLS}    M.~Kr\"amer, E.~Laenen and M.~Spira, Nucl.\ Phys.\ 
                 {\bf B511} (1998) 523
%
\bibitem{cmom}   M. Diemoz, F. Ferroni, E. Longo and G. Martinelli,
                 Z.\ Phys.\ {\bf C39} (1988);\\ 
                 M. Gl\"uck, E. Reya and A. Vogt, Z.\ Phys.\ {\bf C48}
                 (1990) 471; \\
                 Ch.\ Berger, D. Graudenz, M. Hampel and A. Vogt, Z.\
                 Phys.\ {\bf C70} (1996) 77; \\
                 D. A. Kosower, Nucl.\ Phys. {\bf B506} (1997) 439;
                 ibid.\ {\bf B520} (1998) 263 
%
\bibitem{KS95}   A.L. Kataev and V.V. Starshenko, Mod.\ Phys.\ Lett.\
                 {\bf A10} (1995) 235
%
\bibitem{Maple}  D. Redfern, {\it The Maple Handbook (Maple V release 
                 4)}, Springer 1996             
%
\bibitem{MV2}    S. Moch and J.A.M. Vermaseren, Nucl.\ Phys.\ (Proc.\ 
                 Suppl.) {\bf 89} (2000) 131; ibid.~137;\\
                 S. Moch, J.A.M. Vermaseren and M. Zhou, in preparation
%
\bibitem{JB98}   J.~Bl\"umlein and S.~Kurth, Phys.\ Rev.\ {\bf D60} 
                 (1999) 014018
%
\end{thebibliography}
\end{document}